\documentclass[twocolumn]{aastex62}
\usepackage{longtable}
\usepackage{comment}
\usepackage{amsmath, bm, nccmath}
\usepackage{hyperref}

\newcommand{\feh}{\ensuremath{{\rm [Fe/H]}}}
\newcommand{\teff}{\ensuremath{T_{\rm eff}}}

\newcommand{\logg}{\ensuremath{\log{g}}}

\newcommand{\vsini}{\ensuremath{v \sin{i}}}

\newcommand{\rsun}{\ensuremath{{\rm R}_{\odot}}}
\newcommand{\msun}{\ensuremath{{\rm M}_{\odot}}}
\newcommand{\lsun}{\ensuremath{{\rm L}_{\odot}}}

\newcommand{\stname}{TOI-1759}
\newcommand{\stnameTIC}{TIC 408636441}

\newcommand{\rhostar}{\ensuremath{\rho_\star}}
\newcommand{\lmu}{Universit\"ats-Sternwarte, Ludwig-Maximilians-Universit\"at M\"unchen, Scheinerstrasse 1, 81679 M\"unchen, Germany}
\newcommand{\iac}{Instituto de Astrof\'isica de Canarias (IAC), 38205 La Laguna, Tenerife, Spain}
\newcommand{\mpia}{Max-Planck-Institut f\"{u}r Astronomie, K\"{o}nigstuhl  17, 69117 Heidelberg, Germany}
\newcommand{\lsw}{Landessternwarte, Zentrum f\"ur Astronomie der Universit\"at Heidelberg, K\"onigstuhl 12, 69117 Heidelberg, Germany}
\newcommand{\cabesac}{Centro de Astrobiolog\'ia (CSIC-INTA), ESAC, Camino bajo del castillo s/n, 28692 Villanueva de la Ca\~nada, Madrid, Spain}
\newcommand{\ull}{Departamento de Astrof\'isica, Universidad de La Laguna, 38206 La Laguna, Tenerife, Spain}
\newcommand{\iaa}{Instituto de Astrof\'isica de Andaluc\'ia (CSIC), Glorieta de la Astronom\'ia s/n, 18008 Granada, Spain}
\newcommand{\ice}{Institut de Ci\`encies de l'Espai (ICE, CSIC), Campus UAB, c/ de Can Magrans s/n, 08193 Cerdanyola del Vall\`es, Barcelona, Spain}
\newcommand{\ieec}{Institut d'Estudis Espacials de Catalunya (IEEC), c/ Gran Capit\`a 2-4, 08034 Barcelona, Spain}
\newcommand{\iag}{Institut f\"ur Astrophysik, Georg-August-Universit\"at, Friedrich-Hund-Platz 1, 37077 G\"ottingen, Germany}
\newcommand{\hamburg}{Hamburger Sternwarte, Gojenbergsweg 112, 21029 Hamburg, Germany}
\newcommand{\cfa}{Center for Astrophysics \textbar \ Harvard \& Smithsonian, 60 Garden Street, Cambridge, MA 02138, United States of America}
\newcommand{\tls}{Th\"uringer Landessternwarte Tautenburg, Sternwarte 5, 07778 Tautenburg, Germany}
\newcommand{\leuven}{Vereniging Voor Sterrenkunde, Brugge, Belgium \& Centre for mathematical Plasma-Astrophysics, Department of Mathematics, KU Leuven, Celestijnenlaan 200B, 3001 Heverlee, Belgium}
\newcommand{\iris}{AstroLAB IRIS, Provinciaal Domein ``De Palingbeek'', Verbrandemolenstraat 5, 8902 Zillebeke, Ieper, Belgium}
\newcommand{\cabinta}{Centro de Astrobiolog\'ia (CSIC-INTA), Carretera de Ajalvir km 4, 28850 Torrej\'on de Ardoz, Madrid, Spain}
\newcommand{\mig}{Max-Planck-Institute f\"ur Sonnensystemforschung, Justus-von-Liebig-Weg 3, D-37075 G\"ottingen, Germany}
\newcommand{\ames}{NASA Ames Research Center, Moffett Field, CA 94035, United States of America}
\newcommand{\nexsci}{NASA Exoplanet Science Institute/Caltech-IPAC, MC 314-6, 1200 E California Blvd, Pasadena, CA 91125, United States of America}
\newcommand{\goddard}{NASA Goddard Space Flight Center, 8800 Greenbelt Rd., Greenbelt, MD 20771, United States of America}
\newcommand{\ucm}{Departamento de F{\'i}sica de la Tierra y Astrof{\'i}sica \& 
           IPARCOS-UCM (Instituto de F\'{i}sica de Part\'{i}culas y del Cosmos de la UCM), 
           Facultad de Ciencias F{\'i}sicas, Universidad Complutense de Madrid, 28040 Madrid, Spain}
\newcommand{\uofo}{Homer L. Dodge Department of Physics and Astronomy, University of Oklahoma, 440 West Brooks Street, Norman, OK 73019, United States of America}

\newcommand{\mitphysics}{Department of Physics and Kavli Institute for Astrophysics and Space Research, Massachusetts Institute of Technology, Cambridge, MA 02139, United States of America}

\newcommand{\unm}{Department of Physics and Astronomy, University of New Mexico, 210 Yale Blvd NE, Albuquerque, NM 87106, USA}

\newcommand{\seti}{SETI Institute, Mountain View, CA  94043, USA}

\graphicspath{{./}{figures/}}

\received{}
\revised{}
\accepted{}
\submitjournal{AAS}

\shorttitle{TOI-1759~b: a temperate sub-Neptune}
\shortauthors{Espinoza et al.}

\begin{document}

\title{A transiting, temperate mini-Neptune orbiting the M~dwarf TOI-1759 unveiled by \textit{TESS}}

\correspondingauthor{N\'estor Espinoza}
\email{nespinoza@stsci.edu}

\author[0000-0001-9513-1449]{N\'estor Espinoza}
\affiliation{Space Telescope Science Institute, 3700 San Martin Drive, Baltimore, MD 21218, United States of America}

\author[0000-0002-9158-7315]{Enric Pall\'e}
\affiliation{\iac}

\author[0000-0003-3929-1442]{Jonas Kemmer}
\affiliation{\lsw}

\author[0000-0002-4671-2957]{Rafael Luque}
\affiliation{\iaa}

\author[0000-0002-7349-1387]{Jos\'e \,A.~Caballero}
\affiliation{\cabesac}

\author[0000-0003-1715-5087]{Carlos Cifuentes}
\affiliation{\cabesac}

\author{ Enrique Herrero}
\affiliation{\ice}
\affiliation{\ieec}

\author{V\'ictor J. S\'anchez~B\'ejar}
\affiliation{\iac}
\affiliation{\ull}

\author[0000-0002-1166-9338]{Stephan Stock}
\affiliation{\lsw}

\author[0000-0002-0502-0428]{Karan Molaverdikhani}
\affiliation{\lsw}
\affiliation{\mpia}
\affiliation{\lmu} 
\affiliation{ORIGINS: Exzellenzcluster Origins, Boltzmannstraße 2, 85748 Garching, Germany}

\author{Giuseppe Morello}
\affiliation{\iac}
\affiliation{\ull}

\author{Diana Kossakowski}
\affiliation{\mpia}

\author{Martin Schlecker}
\affiliation{\mpia}

% Alphabetical order:
\author{Pedro J. Amado}
\affiliation{\iaa}

\author[0000-0002-0374-8466]{ Paz Bluhm}
\affiliation{\lsw}

\author{Miriam Cort\'es-Contreras}
\affiliation{\cabesac}

\author{Thomas Henning}
\affiliation{\mpia}

\author{Laura Kreidberg}
\affiliation{\mpia}

\author{Martin K\"urster}
\affiliation{\mpia}

\author{Marina Lafarga}
\affiliation{\ice}
\affiliation{\ieec}

\author{Nicolas Lodieu}
\affiliation{\iac}
\affiliation{\ull}

\author{Juan Carlos Morales}
\affiliation{\ice}
\affiliation{\ieec}

\author{Mahmoudreza Oshagh}
\affiliation{\iac}
\affiliation{\ull}

\author{Vera M. Passegger}
\affiliation{\uofo}
\affiliation{\hamburg}

\author{Alexey Pavlov}
\affiliation{\mpia}

\author{ Andreas Quirrenbach}
\affiliation{\lsw}

\author{Sabine Reffert}
\affiliation{\lsw}

\author{ Ansgar Reiners}
\affiliation{\iag}

\author{Ignasi Ribas}
\affiliation{\ice}
\affiliation{\ieec}

\author{Eloy Rodr\'iguez}
\affiliation{\iaa}

\author{Cristina Rodr\'iguez~L\'opez}
\affiliation{\iaa}

\author{Andreas Schweitzer}
\affiliation{\hamburg}

\author{Trifon Trifonov}
\affiliation{\mpia}

% end of alphabetical order
% start of second alphabet order:

\author{Priyanka Chaturvedi}
\affiliation{\tls}

\author{Stefan Dreizler}
\affiliation{\iag}

\author{ Sandra V. Jeffers}
\affiliation{\mig}

\author{Adrian Kaminski}
\affiliation{\lsw}

%mjlg@iaa.es
\author{Mar\'ia Jos\'e L\'opez-Gonz\'alez}
\affiliation{\iaa}

\author{Jorge Lillo-Box}
\affiliation{\cabesac}

\author[0000-0002-7779-238X]{David Montes}
\affiliation{\ucm}

\author{Grzegorz Nowak}
\affiliation{\iac}
\affiliation{\ull}

% pedraz@caha.es
\author{Santos Pedraz}
\affiliation{Centro Astron\'omico Hispano-Alem\'an, Observatorio de Calar Alto,
Sierra de los Filabres, 04550 G\'ergal, Almer\'ia, Spain}

\author{Siegfried Vanaverbeke}
\affiliation{\leuven}
\affiliation{\iris}

\author{Maria R. Zapatero~Osorio}
\affiliation{\cabinta}

\author{Mathias Zechmeister}
\affiliation{\iag}

%%%%%% SG! co-authors (GB photometry): %%%%%
\author{Karen A. Collins}
\affiliation{\cfa}

%ergir@bluewin.ch
\author{Eric Girardin}
\affiliation{Grand Pra Observatory, Switzerland}

%pereguerra@gmail.com
\author{Pere Guerra}
\affiliation{Observatori Astronòmic Albanyà, Camí de Bassegoda s/n, Albanyà 17733, Girona, Spain}

%ramonnavesnogues@gmail.com
\author{Ramon Naves}
\affiliation{Observatori Astronòmic Albanyà, Camí de Bassegoda s/n, Albanyà 17733, Girona, Spain}

%%%%%% Gemini-North co-authors: %%%%%

%i620c788@ku.edu
\author{Ian J.M. Crossfield}
\affiliation{Department of Physics and Astronomy, University of Kansas, Lawrence, KS 66045, United States of America}

\author{Elisabeth C. Matthews}
\affiliation{Observatoire de l’Université de Genève, Chemin Pegasi 51, 1290 Versoix, Switzerland}

\author{Steve B. Howell}
\affiliation{\ames}

\author[0000-0002-5741-3047]{David R. Ciardi}
\affiliation{\nexsci}

\author{Erica Gonzales}
\affiliation{Department of Astronomy and Astrophysics, University of California, Santa Cruz, CA 95060, United States of America}

\author[0000-0001-7233-7508]{Rachel~A.~Matson}
\affiliation{U.S. Naval Observatory, Washington, D.C. 20392, USA}

\author{Charles A. Beichman}
\affiliation{\nexsci}

\author{Joshua E. Schlieder}
\affiliation{\goddard}

%%%%%% S/POC co-authors: %%%%%
\author[0000-0001-7139-2724]{Thomas~Barclay}
\affiliation{University of Maryland, Baltimore County, 1000 Hilltop Circle, Baltimore, MD 21250, USA}
\affiliation{\goddard}

\author{Michael~Vezie}
\affiliation{\mitphysics}

\author{Jesus Noel Villase\~nor}
\affiliation{\mitphysics}

%%%%%% TSO/EXOFOP co-authors: %%%%%
\author[0000-0002-6939-9211]{Tansu Daylan}
\affiliation{\mitphysics}

\author{Ismael Mireies}
\affiliation{\unm}

\author{Diana Dragomir}
\affiliation{\unm}

%%%%% SPOC co author %%%%%%
\author[0000-0002-6778-7552]{Joseph D. Twicken}
\affiliation{\seti}
\affiliation{\ames}

%%%%%% TESS Architects %%%%%%
\author{Jon Jenkins}
\affiliation{\ames}

\author[0000-0002-4265-047X]{Joshua~N.~Winn}
\affiliation{Department of Astrophysical Sciences, Princeton University, 4 Ivy Lane, Princeton, NJ 08544, United States of America}

\author{David Latham}
\affiliation{\cfa}

\author{George Ricker}
\affiliation{\mitphysics}

\author{Sara Seager}
\affiliation{\mitphysics}

\received{}
\revised{}

\begin{abstract}
We report the discovery and characterization of TOI-1759~b, a temperate (400\,K) sub-Neptune-sized exoplanet orbiting the M~dwarf TOI-1759 (TIC~408636441). TOI-1759~b was observed by \textit{TESS} to transit on sectors 16, 17 and 24, with only one transit observed per sector, creating an ambiguity on the orbital period of the planet candidate.
Ground-based photometric observations, combined with radial-velocity measurements obtained with the CARMENES spectrograph, confirm an actual period of $18.85019 \pm 0.00014$\,d. A joint analysis of all available photometry and radial velocities reveal a radius of $3.17 \pm 0.10\,R_\oplus$ and a mass of $10.8 \pm 1.5\,M_\oplus$. Combining this with the stellar properties derived for TOI-1759 ($R_\star = 0.597 \pm 0.015\,R_\odot$; $M_\star = 0.606 \pm 0.020\,M_\odot$; $T_{\textrm{eff}} = 4065 \pm 51$\,K), we compute a transmission spectroscopic metric (TSM) value of over 80 for the planet, making it a {{good}} target for transmission spectroscopy studies.
TOI-1759~b is among the top five temperate, small exoplanets ($T_\textrm{eq} < 500$\,K, $R_p < 4 \,R_\oplus$) with the highest TSM discovered to date.
Two additional signals with periods of 80\,d and {{$>$ 200 d}} seem to be present in our radial velocities. {{While our data suggest {{both could}} arise from stellar activity, the later signal's source and 
periodicity are hard to pinpoint given the $\sim 200$ d baseline of our radial-velocity campaign with CARMENES. Longer 
baseline radial-velocity campaigns should be performed in order to unveil the true nature of this long period signal.}}
\end{abstract}


\section{Introduction} \label{sec:int}

One of the most exciting astronomical developments in the last decade, triggered by 
improved instrumentation and survey designs, is the detection and characterization of small 
($R_p < 4\,R_\oplus$) exoplanets. The study of transiting, relatively low-temperature 
small worlds, in particular, promises to provide key information to understand how different environments 
(e.g. incident stellar fluxes or initial composition) might impact on their bulk properties, and how those might in turn change 
their atmospheric and interior structures \citep{dorn17, nr20,mq2021}. In addition, 
these cooler worlds allow us to make connections with the planets in our own Solar System, 
all of which have equilibrium temperatures smaller than $500$\,K (hereon referred to as ``temperate" 
exoplanets). These connections, in turn, have key 
implications for the search for life outside the Solar System, and have the potential to help us improve 
and refine the concept of planetary habitability itself \citep{tasker17,mb18,seager21}.

Detecting these small, temperate exoplanets is, however, challenging. The relatively longer orbital 
periods needed to have small irradiation levels makes them difficult to detect from the ground 
using the transit technique, which is why most of the known temperate worlds were detected 
by the transit survey with the longest continuous time-baseline: the \textit{Kepler} 
mission \citep{borucki10}. While revolutionary in the search and discovery of small worlds --- 
revealing that they are, in fact, among the most abundant population of 
exoplanets in our galaxy \citep[at least for close-in exoplanets;][]{fp18, hsu19} --- the mission provided few 
systems amenable for further radial-velocity and/or atmospheric characterization, due to the 
inherent faintness of the stars it surveyed. This detailed characterization is fundamental to understand the overall make-up of these small, distant worlds, and help us 
understand and uncover their different sub-populations \citep[see, e.g.][]{zeng19,gupta21, schlichting21,Yu2021}. It is also important to understand fundamental exoplanet demographic questions  such as \textit{why} these small worlds are the most  numerous in our galaxy \citep[see, e.g. ][]{kite19}.

The {Transiting Exoplanet Survey Satellite} \citep[\textit{TESS}]{tess} has been crucial to 
the search for small transiting exoplanets amenable for detailed characterization. To date, it has already doubled the known sample of small, temperate worlds for which masses have been measured 
with follow-up observations. And after 
three years of operation, the mission is just 
starting to exploit its long-time baselines, allowing the discovery of exoplanets on long orbital 
periods. In this work, we present the detection and 
characterization of one such system: TOI-1759~b, an 18.85-day sub-Neptune ($R_p = 3.14\,R_\oplus$, $M_p=10.8\,M_\oplus$), 
orbiting a M~dwarf star.

This paper is structured as follows. In \S~\ref{sec:obs} we describe the data that was obtained 
to understand this new exoplanetary system, which includes photometric, spectroscopic and high-resolution 
imaging data. In \S~\ref{sec:ana} we present the analysis of these data, including the stellar 
and planetary properties of the system. We discuss our results in \S~\ref{sec:dis}, and summarize our 
main conclusions from this work in \S~\ref{sec:conclusions}.

\section{Observations} \label{sec:obs}

\subsection{\textit{TESS}}
\label{sec:tess}

\begin{figure*}
\includegraphics[width=0.7\columnwidth]{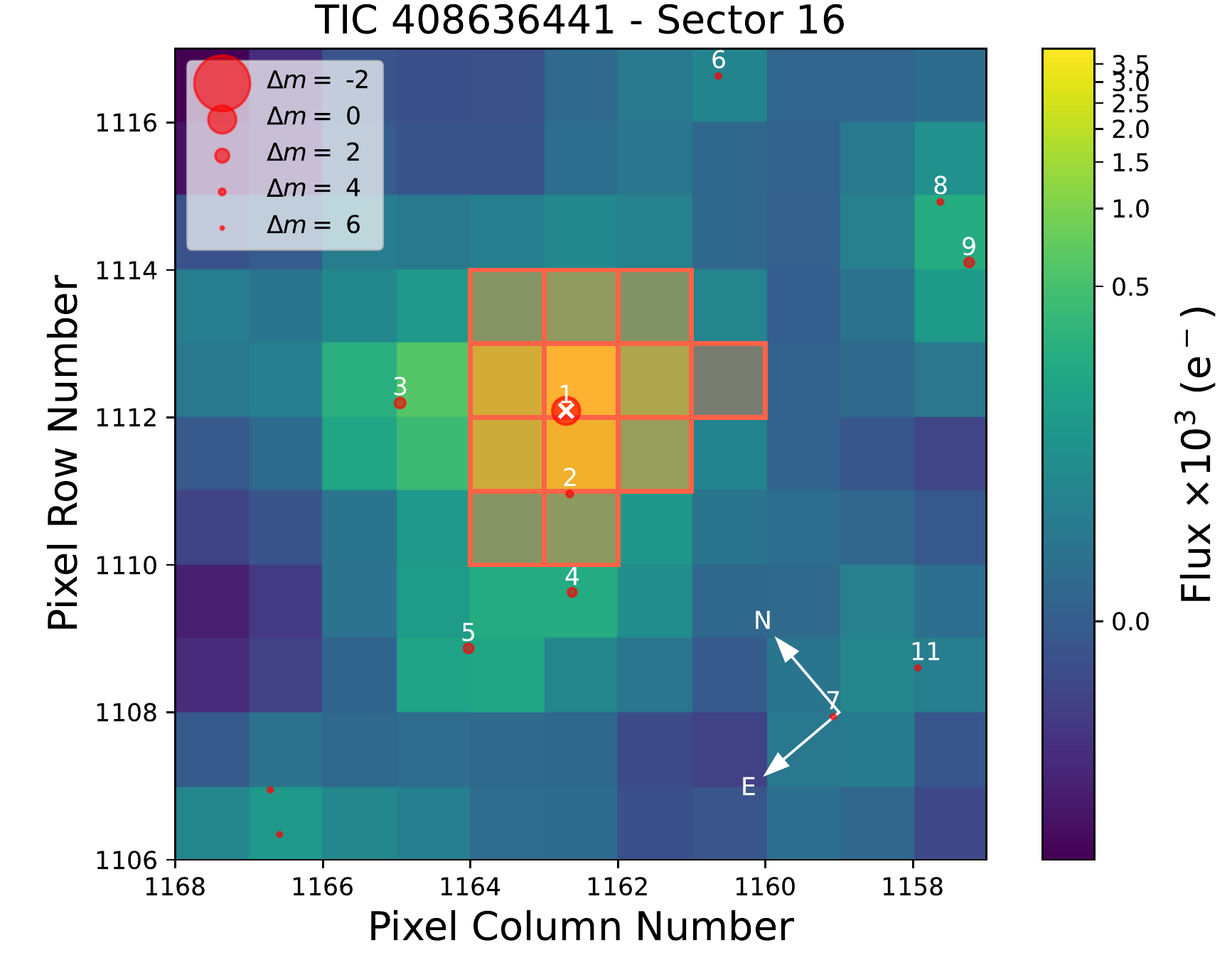}
\includegraphics[width=0.7\columnwidth]{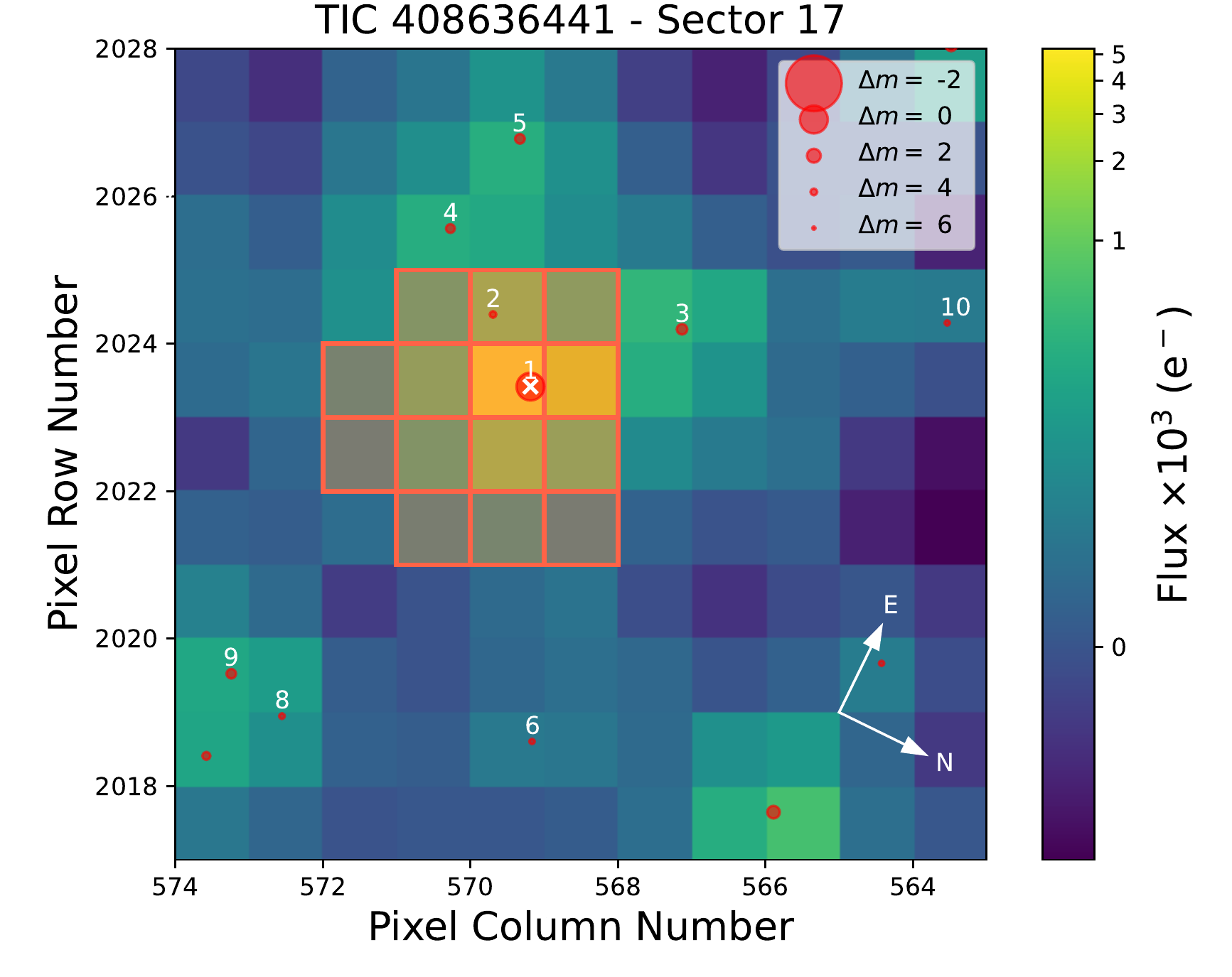}
\includegraphics[width=0.7\columnwidth]{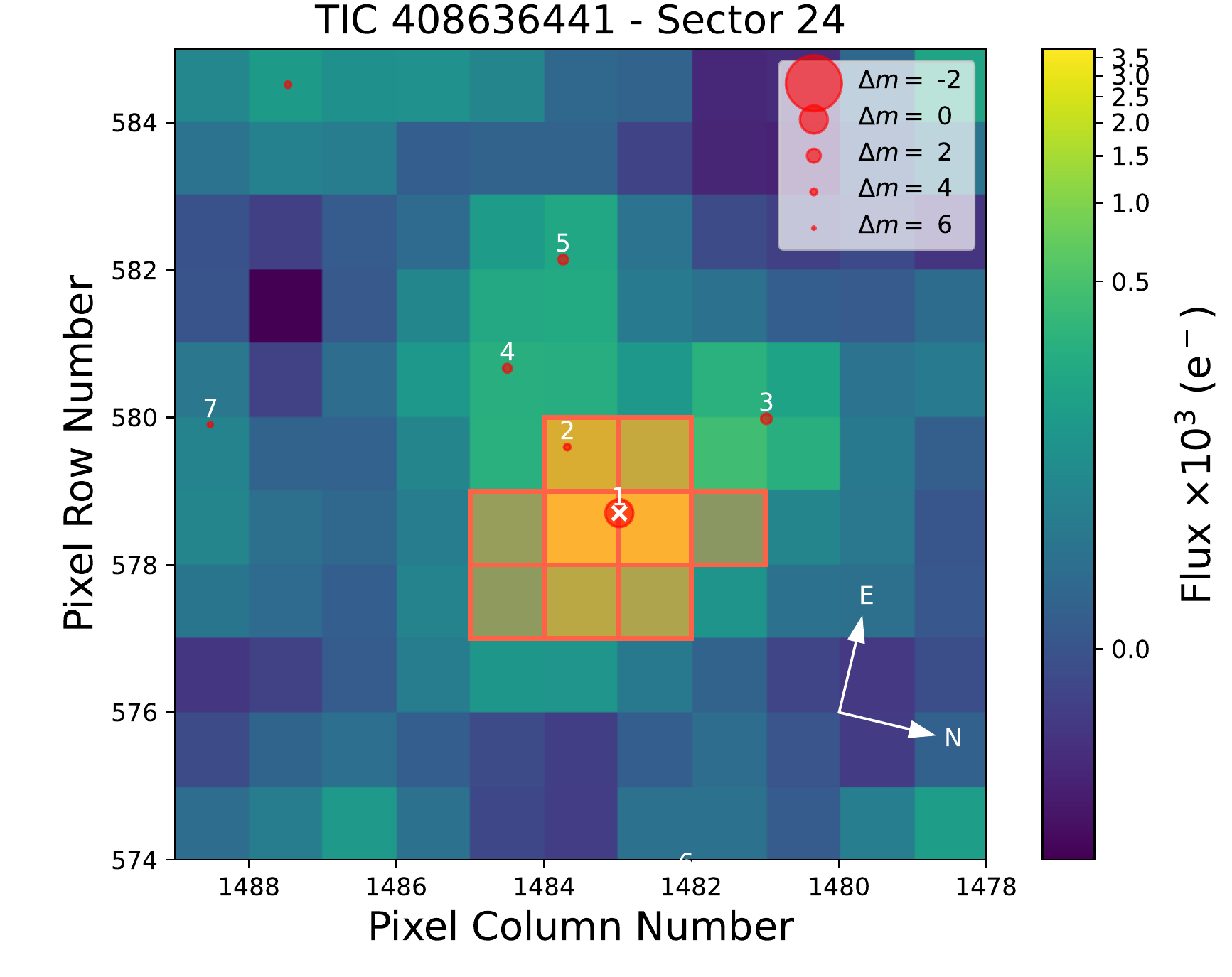}
\caption{\textit{TESS} target pixel files (TPFs) for TOI-1759 from the different 
sectors from which data were gathered with the mission (Sector 16 left; 17 center; 
24, right). TOI-1759 is marked with a white cross on top of a red point, and is numbered as 1. Smaller numbered red points are the closest stars to the target (drawn from {\em Gaia}) with {\em Gaia} magnitude differences with the target of $|\Delta G| < 6$\,mag. Contamination is not a problem for TOI-1759, as most nearby targets are very faint. Plot made using \texttt{tpfplotter} \citep{Aller2020}.
}
\label{fig:tpf}
\end{figure*}

\begin{figure*}
\includegraphics[width=2.1\columnwidth]{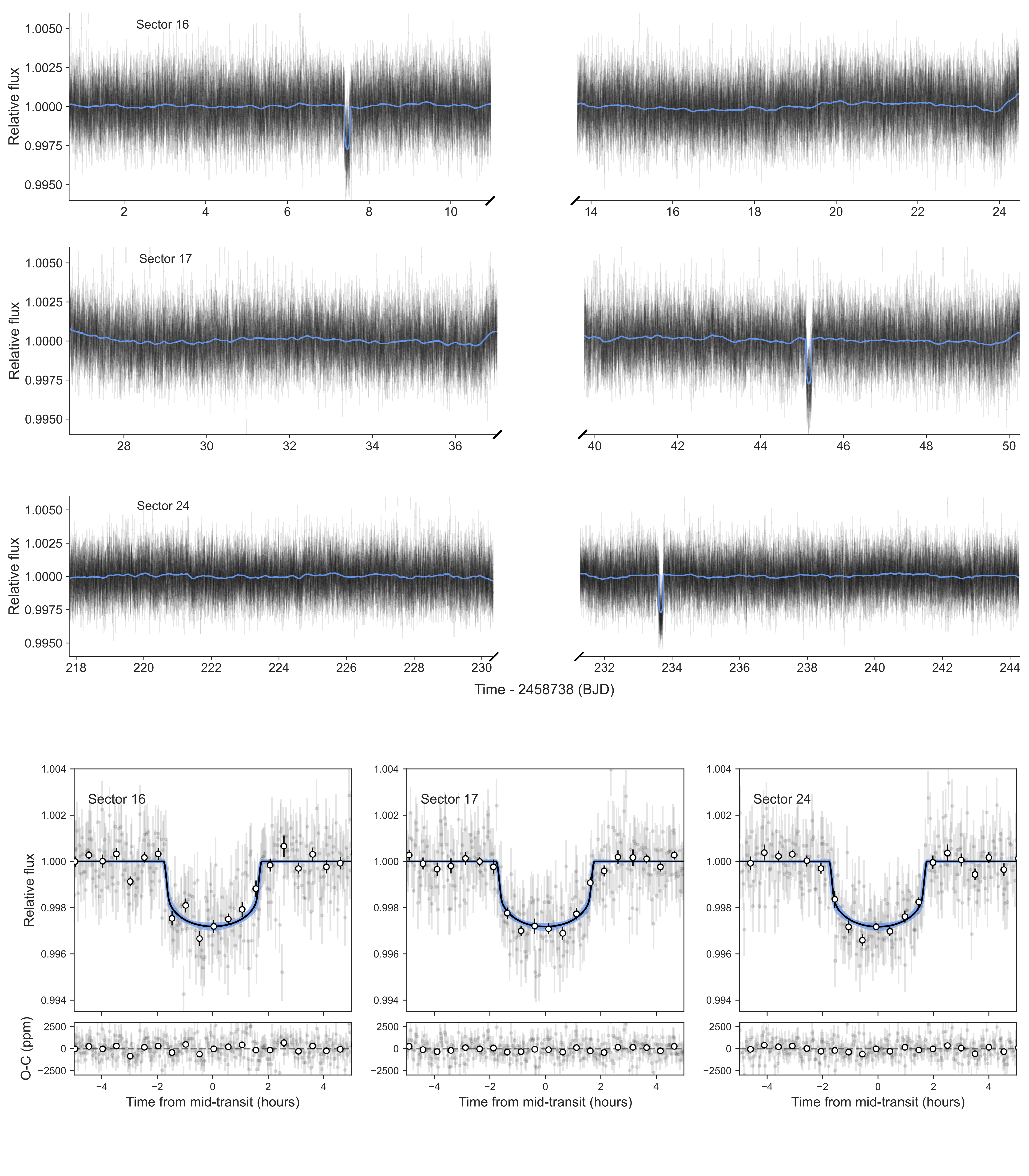}
\caption{\textbf{\textit{TESS} transits of TOI-1759~b}. The top panels present the \textit{TESS} photometry of TOI-1759 in Sectors 16, 17 and 24 as a function of time (black points with errorbars), along with the best-fit model which consists of a transit model plus a Gaussian process (blue curve). Note there is only a single transit observed in each sector. The bottom panels shows a close-up to each of those transits, which have been phased around the time-of-mid transit (grey points with errorbars); the Gaussian process component has been removed from this photometry. The black line in these panels show the best-fit transit model; blue bands represent the 68\% and 95\% credibility bands of the model.}
\label{fig:tess-transits}
\end{figure*}

Observations from \textit{TESS} for TOI-1759 (TYC~4266-736-1, TIC408636441) were obtained during its second year of operation in its high-cadence, 2-minute exposure mode on Sectors 16 (September to October, 2019), 17 
(October to November, 2019) and 24 (April to May, 2020 --- see Figure \ref{fig:tpf} and \ref{fig:tess-transits}{{; the data are also presented in Table \ref{tab:lcs}.}}).

The 2-minute cadence data were processed in the \textit{TESS} Science Processing Operations Center \citep[SPOC;][]{jenkinsSPOC:2016} photometry and transit search pipelines \citep{jenkins2002, jenkins2010} at NASA Ames Research Center. The \textit{TESS} data validation reports \citep{twicken:2018, li:2019} on TOI-1759 \citep{guerrero:2021} show detections of a transiting exoplanet candidate at a 37.7 day period (although the data were also consistent with a planet at half this period, i.e. 18.85 days) and a transit depth of about 2700 ppm. To perform further 
analyses on this target, we 
retrieved the Pre-Data-Conditioning (PDC)-corrected photometry \citep{stumpe:PDC2012,stumpePDC:2014,smithPDC:2012} from all sectors from the Mikulski Archive for Space Telescopes (MAST) archive\footnote{\url{https://archive.stsci.edu/}}, as this is the highest quality 
photometry from the three \textit{TESS} sectors mentioned above. After removing the 
transits of the planet candidate, we ran the Transit Least 
Squares \citep[TLS;][]{tls} algorithm on these photometric time series and found no extra significant signals {{(i.e., signals with a signal-to-noose ratio $>$ 5)}} in the data. 

\subsection{Spectroscopy}

%Table~\ref{tab:rvs}.

\subsubsection{CARMENES}
\label{ssec:spec}

We monitored TOI-1759 with the CARMENES\footnote{Calar Alto high-Resolution search for M~dwarfs 
with Exo-earths with Near-infrared and optical \'Echelle Spectrographs, \url{http://carmenes.caha.es}} instrument located at the 3.5\,m telescope at the Calar Alto Observatory in Almer\'ia, Spain, from July 24, 2020 to January 17, 2021. Our data covered a time span of about 175\,d, over which we were able to detect significant radial-velocity signals. The spectra were processed following the standard CARMENES data flow \citep{Caballero2016} that has been extensively used by previous works \citep[e.g.][]{Zechmeister2018,Morales2019,Trifonov2020}. In our analyses, we only used radial-velocities from the visual (VIS) channel which had mean errors of 2.6 m/s. {{A total of 57 radial-velocity datapoints were used for our analysis, which are presented in Table \ref{tab:rvs}. The spectra used to derive those have a median signal-to-noise ratio of 95 at 840 nm}}. Data from 
our infrared channel was not used as their precision (mean error of 10 m/s) was not enough 
to put meaningful constraints on the radial-velocity variations observed in the VIS channel.

\begin{figure*}
\includegraphics[width=2.2\columnwidth]{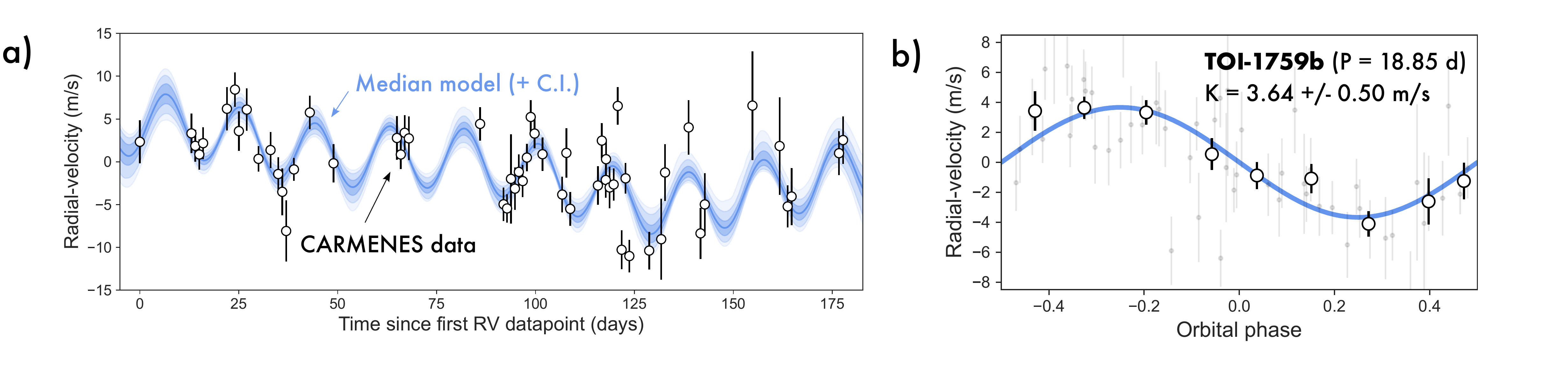}
\caption{\textbf{CARMENES radial-velocity follow-up of TOI-1759}. \textbf{a)} Our CARMENES radial-velocity campaign (white points with black errorbars; {{first point corresponding to }}$2459054.56851$) had a total duration 
of about 6 months, over which we were able to detect both a radial-velocity variation at 
18.85\,d, along with a long-term signal here modeled as a GP. The blue line with transparent bands around it indicate our full (GP + planetary) median signal, along with its 68, 95 and 99\% credibility intervals, respectively. 
\textbf{b)} Phase-folded radial-velocity curve at the period of the transiting exoplanet 
TOI-1759~b ($P=18.85$\,d; grey points, binned datapoints covering about 2\,d plotted as 
white points with black errorbars). The semi-amplitude we obtain for this signal is 
$3.64 \pm 0.50$ m\,s$^{-1}$.}
\label{fig:carmenes-rvs}
\end{figure*}

Figure \ref{fig:carmenes-rvs}a shows 
the radial-velocity as a function of time as observed through the VIS channel, which covers 
the spectral range 520--960\,nm with a spectral resolution of 
$\mathcal{R}$ = 94\,600  \citep{CARMENES, CARMENES18}. Our campaign allowed us to clearly 
detect a signal at about 18.5\,d (consistent with half the period of the transiting exoplanet detected in the \textit{TESS} 
photometry, already discussed in Section \ref{sec:tess}) on top of an additional long-term trend radial-velocity signal. We discuss the details of 
our analysis of these signals in Section \ref{sec:ana}.

\subsection{Ground-based photometry}
\label{ssec:ground}

\begin{figure*}
\includegraphics[width=2.3\columnwidth]{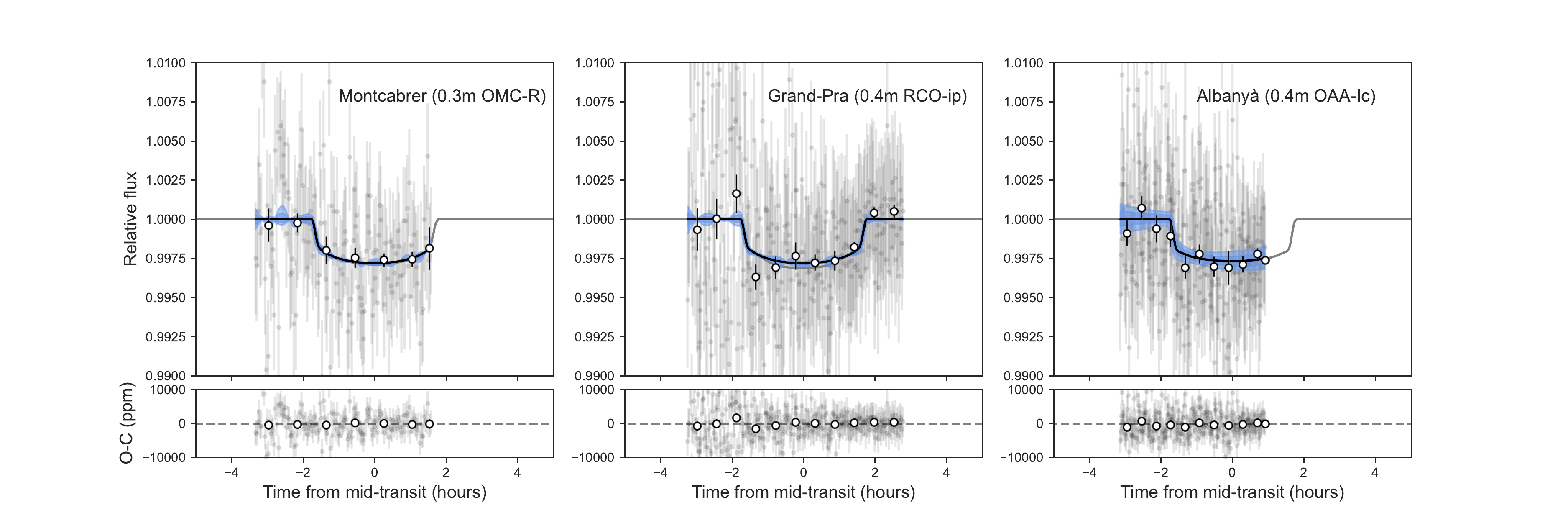}
\caption{\textbf{Ground-based follow-up transit photometry of TOI-1759~b}. Transit of 
TOI-1759~b on May 21, 2020, captured by three different observatories: the OAA telescope 
in the Observatori Astron\`omic Albany\`a, the RCO telescope in the Grand-Pra Observatory 
and the OMC telescope in the Motcabrer Observatori. The duration and depth of the event 
is consistent between instruments.}
\label{fig:gb-transits}
\end{figure*}

Ground-based photometric follow-up observations were performed as part of the \textit{TESS} 
Follow-up Program's (TFOP) Subgroup 1 (SG1). Among the observations, a transit of 
TOI-1759~b in May 21, 2020, was captured by three independent telescopes/observatories: 
the OAA telescope in the Observatori Astron\`omic Albany\`a (Albany\`a, Spain; 4 hours of total observing time, per point precision of 1140 ppm at 1-minute cadence{{; $R$-filter observations}}), the RCO 
telescope in the Grand-Pra observatory (Valais Sion, Switzerland; 6 hours of total observing time, per point precision of 1080 ppm at 1.3-minute cadence{{; $i_p$-filter observations}}), and the OMC telescope 
in the Montcabrer observatory (Barcelona, Spain; 5 hours of total observing time, per point precision of 1500 ppm at 1.9-minute cadence{{; $I_c$-filter observations}}). {{Data reduction for the OAA and OMC observations was performed in a 
two-step process: the \texttt{MaximDL} image processing software was used to perform image calibration 
(bias, darks, flats), while differential photometry was obtained using 
the \texttt{AstroImageJ} software \citep{collins17}. For the RCO observations, image calibration and 
differential photometry were both performed using \texttt{AstroImageJ}.}} The data, along with a best-fit model 
transit after subtracting the best-fit systematics model for each dataset (see Section 
\ref{sec:ana} for details), is presented in Figure \ref{fig:gb-transits}. {{The data are also presented in Table \ref{tab:lcs}.}}

The observed transits by these three independent observatories on May 21, 2020 not only 
confirmed that the event observed by \textit{TESS} was on-target (i.e., happened on TOI-1759), 
but in practice confirmed that the real period of the event was 18.85\,d (i.e., 
half the period proposed by the \textit{TESS} data validation reports), with the 
duration and depth detected by those observatories being consistent with the duration and depth 
observed in the \textit{TESS} transit events.

Long-term photometric monitoring was also performed from the ground using the 0.8\,m Joan Or\'o telescope \citep[TJO;][]{colome10} at the Montsec Observatory in Lleida, Spain {{and the 90-cm telescope at the Sierra Nevada Observatory \citep[SNO;][]{amado:2021}}}. {{For the TJO observations, the data were}} obtained from June 2020 to April 2021, spanning for more than 300\,d and covering 107 different nights. We obtained a total of 1331 images with an exposure time of 40 seconds using the Johnson $R$ filter of the LAIA imager, a 4k×4k CCD with a field of view of 30$'$ and a scale of 0.4$''$/pixel. {{The SNO data were obtained from April to August, 2021, spanning 135 d and 
collecting observations on 55 different nights. Each night, 20 exposures per 
filter were obtained using both Johnson $V$ and $R$ filters, with exposure times of 
60 and 40 seconds, respectively. The photometry from these exposures was 
averaged to obtain a single photometric value per filter each night. These 
data were obtained with a VersArray 2k×2k CCD camera with a field of view of 
13.2x13.2 arcmin$^2$ and a scale of 0.4 arcsec/pixel as well. }} 

{{The TJO CCD images were calibrated with darks, bias and flat fields with the 
ICAT pipeline \citep{colome06}. The differential photometry was extracted 
with \texttt{AstroImageJ} \citep{collins17} using the aperture size that minimized 
the rms of the resulting relative fluxes, and a selection of the 30 brightest 
comparison stars in the field which did not show variability. 
Then, we used our own pipelines to remove outliers and measurements affected 
by poor observing conditions or presenting a low signal-to-noise ratio. This 
resulted in a total of 1087 measurements in the final data set with an 
rms of 6 ppt (parts per thousand). 

In a similar way, the SNO resulting light curves were obtained by the method 
of synthetic aperture photometry. Each CCD frame was also corrected in a 
standard way for bias and flat-fielding. Different aperture sizes were tested
in order to choose the best one for our observations. A number
of nearby and relatively bright stars within the frames were selected as 
reference stars to produce differential photometry of TOI-1759. 
Finally, outliers due to poor observing conditions or very high airmass were 
removed. This resulted in a total of 1029 and 1027 individual data points in 
filters V and R, respectively, with rms of 6.1 and 6.4 ppt. 

Both the TJO and SNO datasets are presented in Table \ref{tab:ltphot}. An analysis 
of these datasets is presented in Section \ref{ssec:rvana}.}}

\subsection{High-resolution imaging}
\label{ssec:imaging}

To help rule out stellar multiplicity and close blends with nearby stars and to obtain more precise planetary radii by accounting for close-in stellar blends \citep{ciardi2015, schlieder2021}, we observed TOI-1759 with both the 'Alopeke\footnote{\url{https://www.gemini.edu/instrumentation/alopeke-zorro}} speckle imaging camera \citep{scott:2018} on the 8\,m Gemini North 
telescope and the NIRC2 near-infrared adaptive-optics fed camera on the 10\,m Keck-II telescope.  The optical and NIR high resolution imaging complement each other with higher resolution in the optical but deeper sensitivity (especially to low-mass stars) in the infrared.

\begin{figure}
\includegraphics[width=\columnwidth]{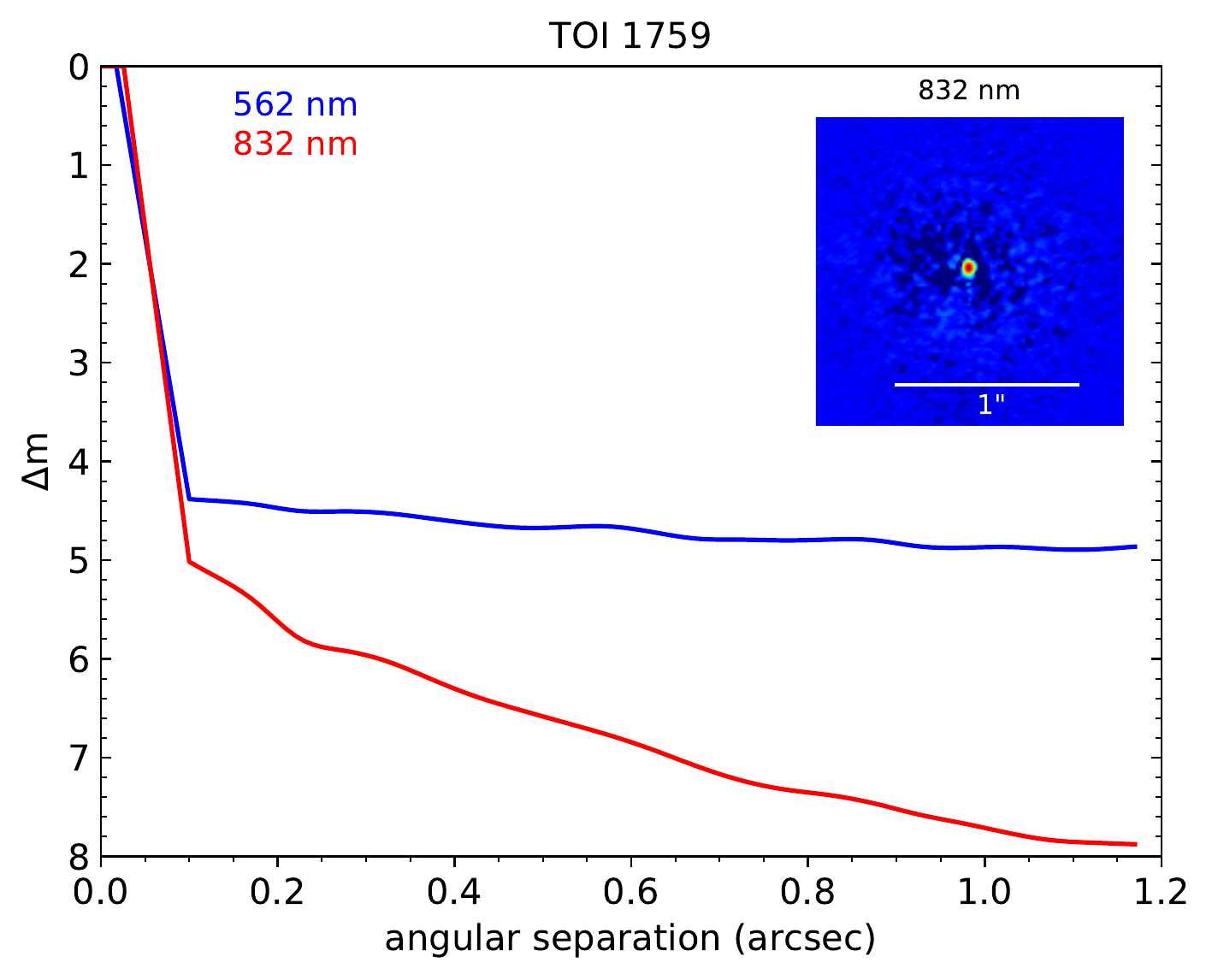}
\caption{\textbf{High spatial resolution imaging of TOI-1759}. Two-band speckle 
imaging observations obtained with the Alopeke speckle imaging camera on the 8\,m Gemini 
North telescope for TOI-1759 reveal no close-companion down to $0.1$'', dimmer than 
about $\Delta m =$ 4-5\,mag than the target.}
\label{fig:imaging}
\end{figure}

'Alopeke obtains diffraction-limited imaging in two simultaneously-imaged narrow 
bands centered at 562 and 832\,nm.  Due to the relative faintness of the target star at 
these wavelengths, we obtained five exposures in each channel, with integration times of 
60\,ms each. We reduced the data using standard techniques using the methods described 
by \cite{matson:2019}.  The resulting contrast curves and reconstructed 832\,nm image, 
all shown in Fig.~\ref{fig:imaging}.  The optical speckle observations show no evidence of an additional stellar companion.

\begin{figure}
\includegraphics[width=\columnwidth]{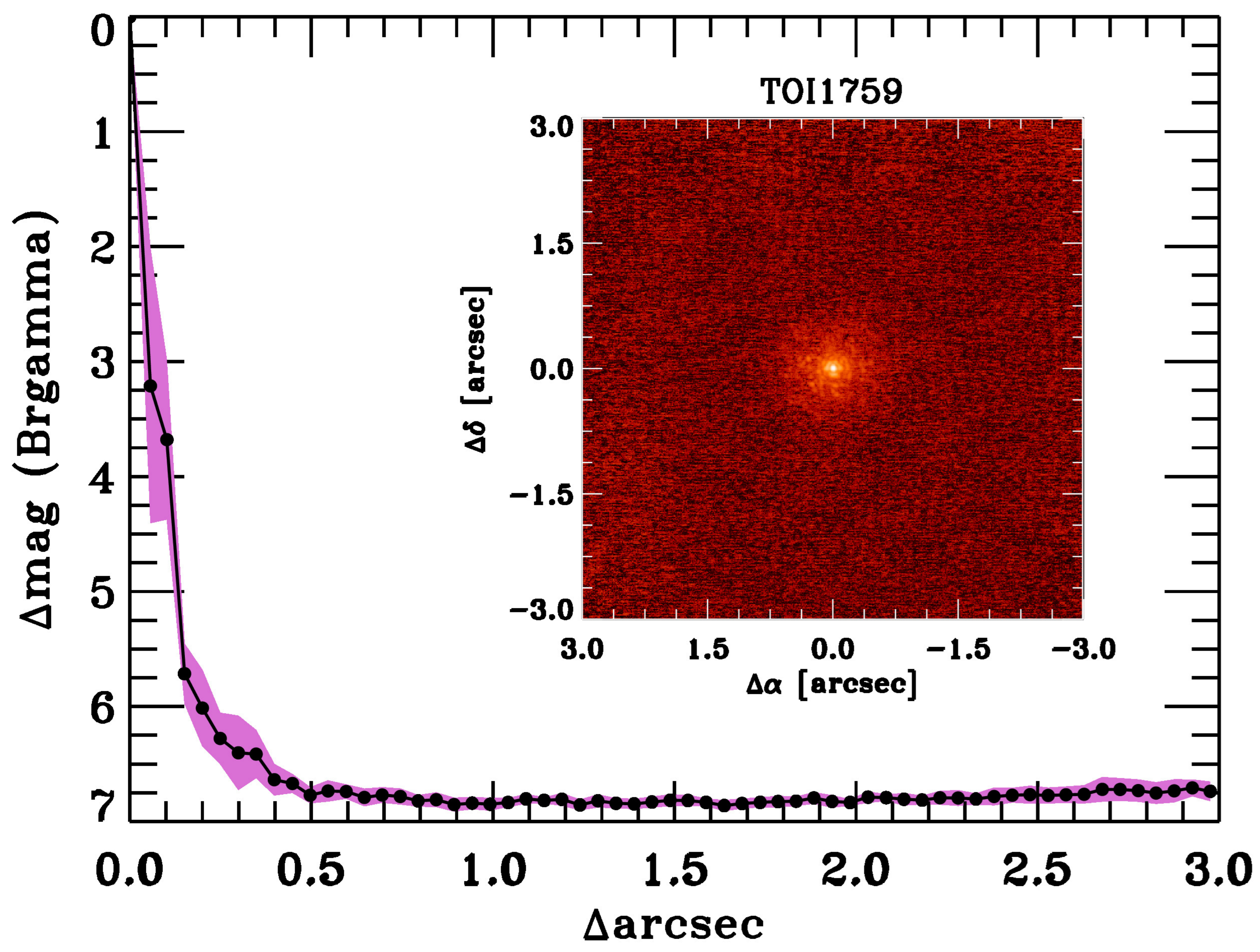}
\caption{\textbf{Near-infrared high spatial resolution imaging of TOI-1759}. NIRC2 
imaging observations obtained for TOI-1759 on Keck-II reveals no close-companion down to $0.1\,\arcsec$, dimmer than about $\Delta m =$ 4-5\,mag than the target.}
\label{fig:imagingao}
\end{figure}

TOI-1759 was also observed with the NIRC2 instrument on Keck-II behind the natural guide star AO system \citep{wizinowich2000}.  The observations were made on 2020~Sep~09 UT in the standard 3-point dither pattern that is used with NIRC2 to avoid the left lower quadrant of the detector which is typically noisier than the other three quadrants. The dither pattern step size was $3\arcsec$ and was repeated twice, with each dither offset from the previous dither by $0.5\arcsec$. The camera was in the narrow-angle mode with a full field of view of $\sim10\arcsec$ and a pixel scale of approximately $0.0099\arcsec$ per pixel. The observations were made in the narrow-band $Br$-$\gamma$ filter $(\lambda_o = 2.1686; \Delta\lambda = 0.0326\mu$m) with an integration time of 1 seconds with one coadd per frame for a total of 9 seconds on target.

The AO data were processed and analyzed with a custom set of IDL tools.  The science frames were flat-fielded and sky-subtracted.  The flat fields were generated from a median average of dark subtracted flats taken on-sky, and the flats were normalized such that the median value of the flats is unity.  Sky frames were generated from the median average of the 9 dithered science frames; each science image was then sky-subtracted and flat-fielded.  The reduced science frames were combined into a single combined image using a intra-pixel interpolation that conserves flux, shifts the individual dithered frames by the appropriate fractional pixels, and median-coadds the frames.  The final resolution of the combined dithers was determined from the full-width half-maximum of the point spread function; 0.049\arcsec.  The sensitivities of the final combined AO image were determined by injecting simulated sources azimuthally around the primary target every $20^\circ $ at separations of integer multiples of the central source's FWHM \citep{furlan2017}. The brightness of each injected source was scaled until standard aperture photometry detected it with $5\sigma $ significance. The resulting brightness of the injected sources relative to the target set the contrast limits at that injection location. The final $5\sigma $ limit at each separation was determined from the average of all of the determined limits at that separation and the uncertainty on the limit was set by the rms dispersion of the azimuthal slices at a given radial distance.  The final combined image and sensitivity curve are shown in Fig.~\ref{fig:imagingao}

Both the optical speckle and the near-infrared adaptive optics observations find no additional stars (down to 0.1'' dimmer than about 4-5 magnitudes than the target in the optical and near-infrared) and so further strengthen the case for TOI-1759~b being a {bona fide} planet.

\section{Analysis} 
\label{sec:ana}

\subsection{Stellar parameters}
\label{ssec:stellarpars}

We obtained the photospheric parameters \teff, \logg\, and \feh\ of TOI-1759 following \cite{Passegger2019} by fitting PHOENIX synthetic spectra to the combined {{(co-added)}} CARMENES VIS spectrum described in Section \ref{ssec:spec}{{, which has a signal-to-noise ratio in the VIS channel of $\approx$ 200}}. {{We used $\vsini =2$\,km\,s$^{-1}$, which was measured by \citet{Marfil2021} as an upper limit.}} We derived its luminosity following \cite{Cifuentes2020} by
using the latest parallactic distance from {\em Gaia} EDR3 \citep{GaiaEDR3}, and by
integrating {\em Gaia}, 2MASS \citep{2MASS} and AllWISE \citep{AllWISE} photometry covering the full spectral energy distribution with the
Virtual Observatory Spectral energy distribution Analyser \citep{Bayo2008}.
The stellar radius follows from Stefan-Boltzmann's law and the stellar mass by using the linear mass-radius relation from \citet{Schweitzer2019}.

TOI-1759's pseudo equivalent width of the H$\alpha$ line as defined
by \citet{Schofer.2019} is  pEW$^\prime({{\rm H}\alpha})<-0.3$\,\AA\ \citep{Marfil2021},
classifying it as an H$\alpha$ inactive star.
Furthermore, \citet{Marfil2021} assign it to the Galactic thin disc population,
which has a maximum age of about 8\,Gyr \citep{Fuhrmann1998}.
Using these two properties as age indicators, we conclude that its age is
between 1\,Gyr (the typical minimum age for field stars) and 8\,Gyr without being able to be more precise.
All collected and derived parameters are presented in Table \ref{tab:stprops}.

\begin{deluxetable*}{lcl}[b!]
\tablecaption{Stellar properties of TOI-1759.  \label{tab:stprops}}
\tablecolumns{3}
\tablewidth{0pt}
\tablehead{
\colhead{Parameter} &
\colhead{Value} &
\colhead{Reference} \\
}
\startdata
Names    &    \stnameTIC  & TIC  \\
 & 2MASS J1472477+6245139 & 2MASS  \\
 & TYC 4266-00736-1 & Tycho-2  \\
 & WISEA J214724.51+624513.8 & AllWISE  \\
    RA  (J2000) &  21$^{\rm h}$47$^{\rm m}$24\fs39 & {\em Gaia} EDR3 \\
    DEC  (J2000) & 62\degr45\arcmin13\farcs7 & {\em Gaia} EDR3  \\
Spectral type & M0.0\,V & Lep13 \\
$\mu_\alpha \cos{\delta}$ [mas yr$^{-1}$] & --173.425 $\pm$ 0.012 & {\em Gaia} EDR3 \\
$\mu_\delta$  [mas yr$^{-1}$] & --10.654 $\pm$ 0.011 & {\em Gaia} EDR3 \\
$\pi$  [mas] &  24.922 $\pm$ 0.010  & {\em Gaia} EDR3 \\ 
$d$ [pc] & 40.112 $\pm$ 0.016 & {\em Gaia} EDR3 \\
\hline
$G_{BP}$   [mag] & 11.7164 $\pm$ 0.0029  & {\em Gaia} EDR3\\
$G$    [mag] & 10.8386 $\pm$ 0.0028 & {\em Gaia} EDR3\\
$T$ [mag]  & 9.9284 $\pm$ 0.0073 & TIC\\
$G_{RP}$   [mag] & 9.9174 $\pm$ 0.0038 & {\em Gaia} EDR3\\
$J$   [mag] & 8.771 $\pm$ 0.043 & 2MASS\\
$H$   [mag] & 8.114 $\pm$ 0.059 & 2MASS\\
$K_s$   [mag] & 7.930 $\pm$ 0.020 & 2MASS\\
$W1$   [mag] & 7.825 $\pm$ 0.027 & AllWISE\\
$W2$   [mag] & 7.886 $\pm$ 0.020 & AllWISE\\
$W3$   [mag] & 7.787 $\pm$ 0.018 & AllWISE\\
$W4$   [mag] & 7.643 $\pm$ 0.111 & AllWISE\\
\hline
$L_\star$  [10$^{-4}$\,\lsun] & 876.7 $\pm$ 6.3 & This work \\
\teff  \, [K] & 4065 $\pm$ 51 & This work \\
\logg \, [dex] & 4.65 $\pm$ 0.04 & This work \\
\feh  \, [dex] & 0.05 $\pm$ 0.16 & This work \\
\vsini  \,[km s$^{-1}$] & $\le$ 2 & Mar21 \\
$M_\star$  \,[\msun] & 0.606 $\pm$ 0.020 & This work\\
$R_\star$ \,[\rsun] & 0.597 $\pm$ 0.015 & This work \\
Age  [Gyr] & 1--8 & This work\\
\rhostar \, [kg m$^{-3}$] & 3949 $\pm$ 323 & This work\\
\enddata
\tablerefs{2MASS \citep{2MASS}, AllWISE \citep{AllWISE}, {\em Gaia} EDR3 \citep{GaiaEDR3}, Lep13 \citep{Lepine2013}, Mar21 \citep{Marfil2021}, TIC \citep{Stassun2019}, Tycho-2 \citep{Hog2000}. }
\end{deluxetable*}

\subsection{Radial-velocity analysis}
\label{ssec:rvana}

\begin{figure*}
\includegraphics[width=2.15\columnwidth]{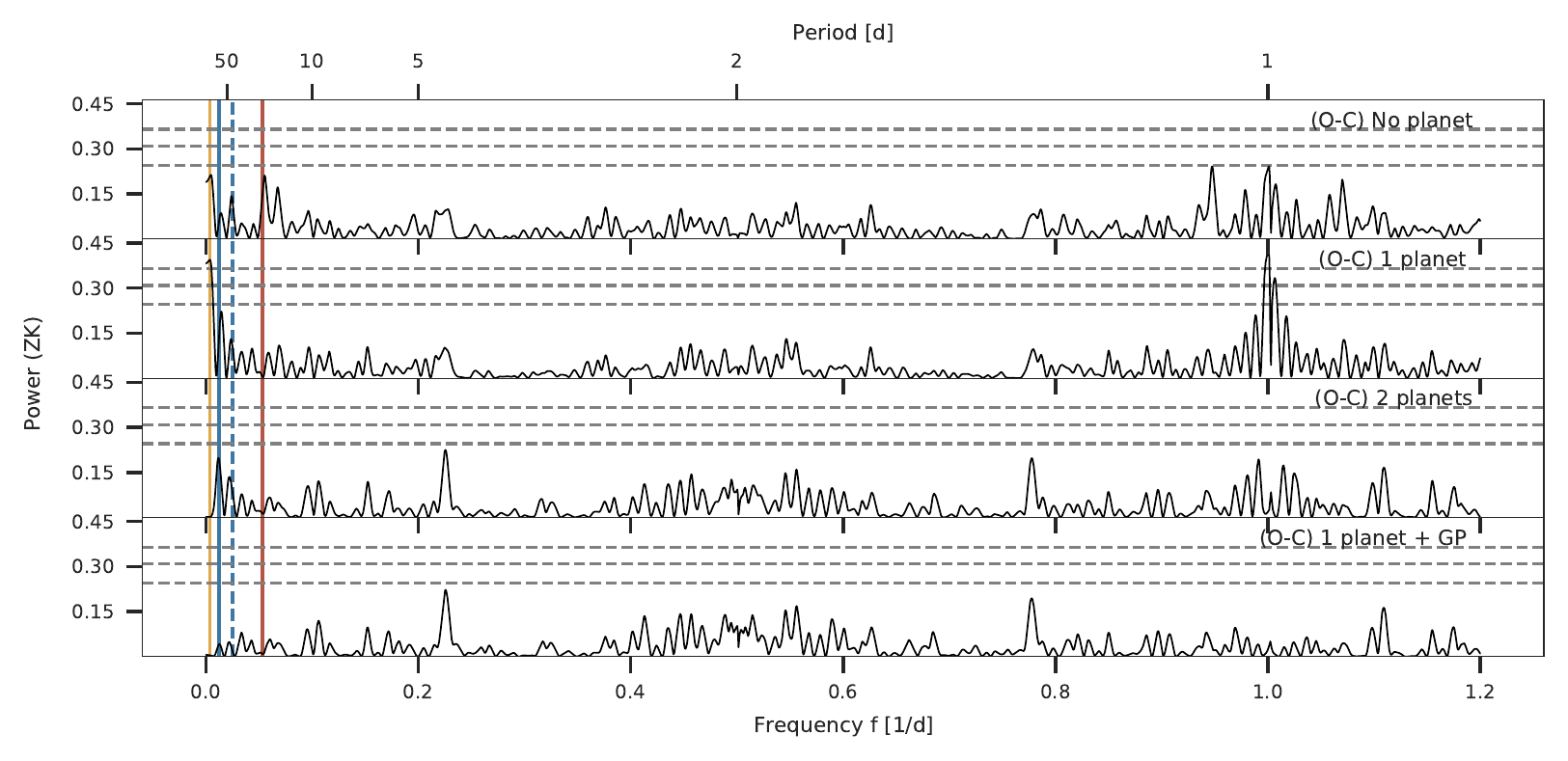}
\caption{\textbf{Generalized Lomb-Scargle (GLS) periodograms of the CARMENES radial-velocity data after subtracting different models}. The power at each 
frequency corresponds to the one defined in \citet[][ZK in the labels above]{zk:2009}. The period of the transiting planet, $P=18.85$\,d, is marked by the red solid line and the stellar rotation period, $P\approx80$\,d, and its first harmonic with the blue solid and dashed lines respectively. Further, the additional present long term periodicity with $P\approx270$\,d is marked by the yellow solid line. The horizontal dashed grey lines show the analytical false alarm probabilities (FAP)
of 10, 1, and 0.1\%. {{The top panel corresponds to the mean-substracted CARMENES radial-velocity dataset.}}}
\label{fig:gls_rv}
\end{figure*}

We performed a detailed analysis on the radial-velocities described in Section 
\ref{ssec:spec} in order to constrain the possible signals arising from these data. 
To this end, we performed a suite of model fits to the radial-velocity data using 
\texttt{juliet} \citep{espinoza:juliet}, in order to measure the evidence for a 
planet in the data using Bayesian evidences, $Z = P(\textnormal{Model} | \textnormal{Data})$. 
The fits were performed using the Dynamic Nested Sampling algorithm implemented in the \texttt{dynesty} 
library \citep{dynesty}.

%The radial-velocity models we considered consisted mainly of three classes of models. 

We considered three main types of radial velocity models.
The first was a ``no planet" model, namely, a set of models in which it is assumed there is no planetary signal present in the radial-velocity data, and 
which thus assumes the data are either consistent with a flat line or with correlated noise modeled through a 
Gaussian process (GP). The second class were 
``1-planet" models; these considered the presence of a planetary signal in the 
radial-velocity data (modeled as a circular orbit), and a suite of possible extra signals, such as linear or quadratic trends, or a (quasi-periodic) GP. 
Finally, we also considered the possibility that the data was best explained by a 
``2-planet" model, as a sum of two circular orbits and a suite of possible extra signals, 
such as a linear, quadratic or a GP trend. 

We first performed ``blind" fits to the data --- that is, fits in which we assumed no 
strong prior knowledge on the signal(s) present on our radial-velocities. For our GP, 
we assumed a quasi-periodic kernel of the form
\begin{equation*}
    k(t_i, t_j) = \sigma^2_{GP} \exp \left( -\alpha \tau^2 - \Gamma \sin^2 \left[\frac{\pi \tau}{P_{rot}} \right] \right),
\end{equation*}
where $\tau = |t_i - t_j|$. We set log-uniform priors for $\sigma_{GP}$, $\alpha$ and 
$\Gamma$, with lower and upper limits of $(0.01, 100)$\,m\,s$^{-1}$, $(10^{-10}, 1)$\,d$^{-1}$ and 
$(0.01, 100)$ respectively, based on the experiments performed with this kernel in \cite{stock2020a} 
and \cite{stock2020b}, and a uniform prior for $P_{rot}$ between 0.5 (half the best 
sampling in our radial-velocities) and 350\,d (two-times our time-baseline). 
The linear and quadratic trends both had uniform priors on the coefficients of 
$(10^{-3}, 10^3)$. As for the circular orbits, we set a uniform prior on the period of the first one 
from $0.5$ to 50\,d (so as to cover the 18 and 36-day periods which could possibly originate 
from the transiting exoplanet), and a uniform prior on the period of the second one from 50 to 
350\,d. Uniform priors 
were set for the {{time of inferior conjunction}} for both covering the entire time-baseline of our 
observations, the semi-amplitude --- between 0 and 100\,m\,s$^{-1}$ --- and the systematic 
radial velocity --- between -100 and 100\,m\,s$^{-1}$. A jitter term, $\sigma_w$ was added to all our fits with a log-uniform prior between 0.01 and 100 \,m\,s$^{-1}$. 

In our ``blind" fits, we found that all models considering a periodic component were consistent with a prominent signal at $\sim 18.5$\,d, 
which corresponds to the transit signal implied by the \textit{TESS} photometry presented in Section \ref{sec:tess} and the 
ground-based transits presented in Section \ref{ssec:ground}. The model with the highest evidence in our set of fits was one composed of a sinusoid  plus a quasi-periodic GP ($|\Delta \log Z| \approx 4.3 $, compared with the no-planets model). Given the high-resolution imaging data presented in Section \ref{ssec:imaging}, the ground-based transit detected on-target presented in \ref{ssec:ground} and the fact that the period 
of {{the planetary signal for}} the model with the highest evidence {{($18.49^{+0.23}_{-0.21}$ d) agrees with}} the period 
implied by the photometric data {{($18.8480 \pm 0.0010$ d)}}, we consider that this 18-day period signal in both photometry and radial-velocities 
is, indeed, a {bona fide} transiting exoplanet.

% In our ``blind" fits, the model with the highest evidence 
% in our set of fits was one composed of two simple Keplerians, although this model was 
% indistinguishable ($|\Delta \log Z| < 2$) with the no planets model. Interestingly, however, 
% the periods retrieved by the two-planet model were $18.46^{+0.21}_{-0.19}$\,d and 
% $261^{+45}_{43}$\,d; the first period being consistent with the 18-day period transit signal 
% implied by the \textit{TESS} photometry presented in Section \ref{sec:tess} and the ground-based 
% transits presented in Section \ref{ssec:ground}. While the model with only two 
% Keplerians has a log-evidence difference with a flat-line model (composed of only a systematic 
% radial-velocity component and a jitter 
% term) of only $\Delta \log Z = 1.31$ --- i.e., the two-planet model is 3 times more likely given the data 
% than the flat line --- given the high-resolution imaging data presented in Section \ref{ssec:imaging}, the 
% ground-based transit detected on-target presented in \ref{ssec:ground} and the fact that one of the periods 
% of the model with the highest evidence in this blind fit perfectly matches the period implied by the 
% photometric data, we consider that this 18-day period signal in both photometry and radial-velocities 
% is, indeed, a {\em bona-fide} transiting exoplanet.

Having concluded that the 18-day period signal is indeed a {bona fide} transiting 
exoplanet, we then focused on finding the best model that explains the radial-velocity 
dataset. We considered the same class of models and priors as the ones presented above, 
but now for the first planet we fixed the period and transit center to the values defined by a photometric fit made to the data using \texttt{juliet} (see 
Section \ref{sec:glob} for details on the priors of that fit): period $P=18.85008 \pm 0.00018$\,d, and time-of-transit center $t_0 = 2458745.4651 \pm 0.0015$\,d. A compilation of the log-evidences for each of the fits we performed is presented in 
Table \ref{tab:logZ}. As can be seen, the model with the highest evidence is 
once again a 1-planet + GP model. Interestingly, however, this model is in practice 
indistinguishable \citep[$|\Delta \log Z| < 2$;][]{t2008} from most of the 2-planet models (except the 2-planet + 
linear trend model). It is also indistinguishable from all those models considering either 
one or both of them having eccentric orbits.

\begin{deluxetable}{lr}[b!]
\tablecaption{Log-evidence differences $\Delta Z$ between different models considered for 
our radial-velocity-only analysis, assuming one of the Keplerian signals to have the 
same ephemerides as those implied by the observed transits. 
Below, a flat-line model includes only a systematic radial-velocity and a jitter term. The GP 
refers to a Gaussian Process with a quasi-periodic kernel. See text for details on the 
priors. \label{tab:logZ}}
\tablecolumns{2}
%\tablenum{2}
\tablewidth{10pt}
\tablehead{
\colhead{Model} &
\colhead{$\ln \Delta Z$}
}
\startdata
\multicolumn{2}{l}{\textit{No planet models}} \\
%\vspace{0.1cm}
\ \ \ \ Flat line & --9.4 \\
\ \ \ \ GP & --6.6 \\
\multicolumn{2}{l}{\textit{1 planet models}} \\
%\vspace{0.1cm}
\ \ \ \ 1 planet + linear trend & --23.6 \\
\ \ \ \ 1 planet + quadratic trend \ \ \ \ \ \ \ \ \ \ \ \ \ \ \ \ \ \  \ \ \ \ \ \ \ \ \ & --9.1 \\
\ \ \ \ 1 planet & --8.8 \\
\ \ \ \ 1 planet + GP & 0 \\
\multicolumn{2}{l}{\textit{2 planet models}} \\
%\vspace{0.1cm}
\ \ \ \ 2 planet + linear trend & --12.6 \\
\ \ \ \ 2 planet & --1.6 \\
\ \ \ \ 2 planet + GP & --1.4 \\
\ \ \ \ 2 planet + quadratic trend & --1.3 \\
%r1\tablenotemark{a} & $U(0,1)$ & \rone \\
%r2\tablenotemark{a} & $U(0,1)$ & \rtwo \\
\enddata
\end{deluxetable}

It is interesting to note that the posterior distribution function of the GP rotation period of the 
1-planet + GP model, $P_\textrm{rot}$, shows a bimodal distribution, with peaks at $\sim80$\,d and $>150$\,d. In the 2-planet + GP model, the latter long periodic signal is picked up by the second sinusoid and is constrained to be about $270 \pm 60$\,d, while the quasi-periodic component of the GP shows again an accumulation of samples with higher likelihood at $\sim80$\,d. A Generalized Lomb-Scargle 
\citep[GLS,][]{zk:2009} periodogram analysis of the residuals of each of those models is presented 
in \autoref{fig:gls_rv}, in order to further explore the nature of these long-period signals. 
If the signal of the transiting planet is subtracted (second panel), a significant {{power excess}} is apparent at $\sim270$\,d {{(which 
is above the baseline of our observations, which was 175 d)}} and next to it, even though insignificant, a peak at $\sim80$\,d. The latter is independent of the long term variation, as it is still present after fitting the transiting planet together with the $\sim270$\,d signal (third panel).

\begin{figure}
\includegraphics[width=1\columnwidth]{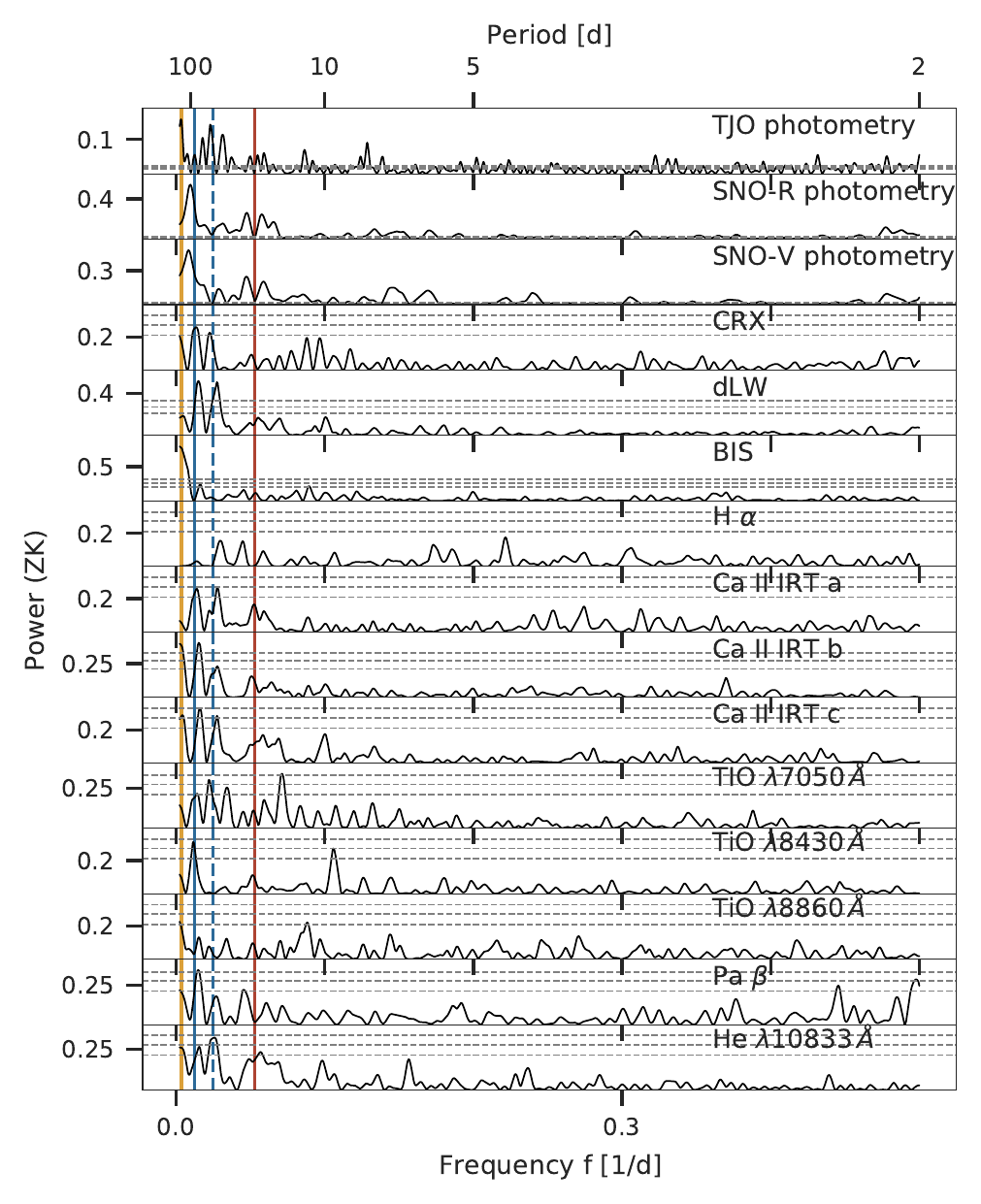}
\caption{\textbf{GLS periodograms of the TJO photometry{{, SNO photometry}} and the CARMENES spectral activity indicators}. Analogous to the periodograms of the RVs, the period of the transiting planet, $P=18.85$\,d and the additional present long term periodicity with $P\approx270$\,d are marked  by the solid {{red and yellow}} lines, respectively. The stellar rotation period, $P\approx80$\,d, and its first harmonic are indicated by the blue solid and dashed lines. The horizontal dashed grey lines show the analytical false alarm probabilities (FAP)
of 10, 1, and 0.1\%.}
\label{fig:gls_activity}
\end{figure}

To find out whether the $\sim80$\,d {{signal}} or {{the}} $\sim270$\,d {{power excess}} could be attributed to stellar activity, we investigated the activity indicators that are routinely derived from the CARMENES spectra \citep[see][for the full list of indicators and how they are calculated]{Zechmeister2018,Schofer.2019,Lafarga.2020}. In \autoref{fig:gls_activity} we show the GLS periodograms of {{several activity indicators}} and the long term photometry presented in Section \ref{ssec:ground}. Almost all indicators, as well as the photometry, show consistent signals around 80 and/or 40\,d, which would explain the $\sim80$\,d seen in the RVs as the stellar rotation period. {{If indeed the rotation period of the star is 80 days, the 40-day signal could be interpreted as spots at 
opposite longitudes and/or a by-product of the not-strictly-periodic nature of the signal}}. Further, the TJO data, the BIS and the Ca~{\sc ii} IRT b and c also show a long term trend, which might be related with the $\sim 270$\,d {{power excess}}.

Given that we could not rule out a stellar origin for the $\sim80$\,d {{signal}} and {{the}} $\sim270$\,d {{power excess,}} and that the 1-planet + GP model has the highest evidence for our RV data and can account for all significant signals in the data (see the residuals in the last panel of \autoref{fig:gls_rv}), we consider this 
model for the global modelling of the data which we present in the next subsection. We note that we also tested fitting our global model using the rest of the models in Table \ref{tab:logZ} which are indistinguishable to the one being selected here, and all of them gave rise to similar constraints in the final parameters of the transiting exoplanet.

% As can be observed from Table \ref{tab:logZ}, the model with the highest evidence is, once 
% again, a 2-planet model. Interestingly, however, this model is in practice 
% indistinguishable ($|\Delta \log Z| < 2$) from that of a 2-planet model plus a GP and that of 
% a 1-planet model plus a GP. It is also indistinguishable from all those models considering either 
% one or both of them having eccentric orbits. The second Keplerian in the 2-planet fit has a 
% period that ranges from 200\,d to 1000\,d, with a peak at about 500\,d. The same peak is found in the 2-planet plus GP fit, where the quasi-periodic kernel period is constrained to be about 
% $P_{rot}$ of 74\,d. The same component in the 1-planet plus GP fit has a period of 32\,d. 
% {\textcolor{red}{\textbf{Jonas}: I think a nice plot of the periodogram of both the RVs and the 
% activity indicators will be good here. Basically, I think it's interesting we recover a 
% Prot of 74\,d on the 2-planet + GP fit, as we do see some activity peaks close to 70\,d 
% in the activity indicators.}}

% Given the results in Table \ref{tab:logZ}, we move forward considering the 2-planet model 
% as the model that best explains our current radial-velocity data.

\subsection{Global modeling}
\label{sec:glob}

We performed a global modelling of the radial-velocity and photometric data using the 
\texttt{juliet} library \citep{espinoza:juliet} in order to jointly constrain the planetary 
properties from the photometry and radial-velocity datasets outlined in previous 
sections. As in the radial-velocity analysis presented in \ref{ssec:rvana}, we once again 
use the Dynamic Nested Sampling algorithm implemented in the \texttt{dynesty} library 
\citep{dynesty}.

For the \textit{TESS} photometry, we decided to use a GP to consider residual systematic 
trends in the PDC lightcurves under use in this work. We used an Exponential-Mat\'ern 
kernel (i.e., the product of an exponential and a Mat\'ern 3/2 kernel) as implemented in 
the \texttt{celerite} library \citep{celerite} via \texttt{juliet}, with hyperparameters (GP amplitude, $\sigma_{GP}$, and two time-scales: one for the Mat\`ern 3/2 part of the kernel, $\rho$, and 
another for the exponential part of the kernel, $T$) 
which are individual to each of the sectors; a jitter term is also added in quadrature to 
the covariance matrix for each sector. A quadratic law is used to constrain limb-darkening, where the 
coefficients are shared between the different sectors; we use the parametrization of 
\cite{Kipping:LDs} instead of fitting for the limb-darkening coefficients directly. For the 
ground-based photometry, we found that airmass was a very good predictor of the long-term 
trends in the data, and so we added this as a linear regressor in our fit --- weighted by 
a coefficient $\theta$, which is different for each instrument and is jointly 
fit with the rest of the parameters of the global fit. In addition, we observed that 
this linear regressor was insufficient to model all the correlated noise leftover on the OMC 
and RCO datasets. We thus decided to fit those with an additional Mat\`ern 3/2 kernel. A linear 
limb-darkening law was used for all ground-based instruments, as a higher order law 
was not necessary given the lower photometric precision \citep[see, e.g. ][]{espinoza:2016:lds}. An individual jitter term was added to the diagonal of 
the covariance matrix on each of to those datasets as well. For the radial-velocities, following our results in Section \ref{ssec:rvana}, we consider 
a 1-planet model plus a quasi-periodic GP as the model to be fit in our joint analysis. We set a wide prior for the systemic 
radial-velocity, as well as for the jitter term and the hyperparameters of the 
GP --- in particular, for the period of the quasi-periodic GP $P_\textrm{rot}$, we use a 
wide period between 20 and 350\,d in order to cover the two possible periods for this 
parameter observed in Section \ref{ssec:rvana}. The full definition of the priors and 
corresponding posteriors of our joint fit are given in Table \ref{tab:plprops}.

\startlongtable
\begin{deluxetable*}{lcc}
\tablecaption{Prior and posterior parameters of the global fit performed to TOI-1759. 
For the priors, $N(\mu,\sigma^2)$ stands for a normal distribution with mean $\mu$ 
and variance $\sigma^2$, $TN(\mu,\sigma^2; a, b)$ is a truncated normal distribution with lower and 
upper limits given by $a$ and $b$, respectively; $U(a,b)$ and $\log U(a,b)$ stands for a uniform and log-uniform distribution between $a$ and $b$, respectively.\label{tab:plprops}}
\tablecolumns{3}
%\tablenum{2}
\tablewidth{0pt}
\tablehead{
\colhead{Parameter} &
\colhead{Prior} &
\colhead{Posterior} \\
}
\startdata
\multicolumn{3}{l}{\textit{Stellar \& planetary parameters}} \\
\ \ \ \ $P_1$ [d] & $N(18.85,0.1^2)$  & $18.85019 \pm 0.00013$ \\
\ \ \ \ $t_{0,1}$ (BJD) &  $N(2458745.45,0.1^2)$ &  $2458745.4654 \pm 0.0011$ \\
\ \ \ \ $R_{p,1}/R_\star$  &  $U(0.0,1.0)$ &  $0.0483 \pm 0.0010$ \\
\ \ \ \ $b_1=(a/R_\star)\cos(i)$  &  $U(0.0,1.0)$ &  $0.21^{+0.09}_{-0.10}$ \\
\ \ \ \ $K_1$ [m\,s$^{-1}$] &  $U(0,100)$ &  $3.64^{+0.50}_{-0.51}$ \\
\ \ \ \ $e_1$  &  fixed &  0 \\
\ \ \ \ $\omega_1$  &  fixed &  90 \\
\ \ \ \ $\rho_\star$ [kg\,m$^{-3}$] &  $TN(3949,323^2; 1000, 10000)$ &  $3970^{+218}_{-233}$ \\
\multicolumn{3}{l}{\textit{TESS photometry instrumental parameters}} \\
\ \ \ \ $q_{1,TESS}^a$ &  $U(0,1)$ &  $0.32^{+0.24}_{-0.14}$ \\
\ \ \ \ $q_{2,TESS}^a$ &  $U(0,1)$ &  $0.46^{+0.28}_{-0.25}$ \\
\ \ \ \ $m_\textnormal{flux,16}$ [ppm] &  $U(0,10^5)$ &  $851^{+7916}_{-7770}$ \\
\ \ \ \ $m_\textnormal{flux,17}$ [ppm] &  $U(0,10^5)$ &  $-1598^{+8062}_{-8089}$ \\
\ \ \ \ $m_\textnormal{flux,24}$ [ppm] &  $U(0,10^5)$ &  $428^{+7904}_{-7797}$ \\
\ \ \ \ $\sigma_\textnormal{w,16}$ [ppm] &  $\log U(0.1,10^5)$ &  $1.8^{+7.5}_{-1.5}$ \\
\ \ \ \ $\sigma_\textnormal{w,17}$ [ppm] &  $\log U(0.1,10^5)$ &  $2.9^{+15.1}_{-2.5}$ \\
\ \ \ \ $\sigma_\textnormal{w,24}$ [ppm] &  $\log U(0.1,10^5)$ &  $1.9^{+7.7}_{-1.6}$ \\
\ \ \ \ $\sigma_\textnormal{GP,16}$ [ppm] &  $\log U(10^{-4},10^2)$ &  $0.000104^{+0.000047}_{-0.0000026}$ \\
\ \ \ \ $\sigma_\textnormal{GP,17}$ [ppm] &  $\log U(10^{-4},10^2)$ &  $0.000106^{+0.000067}_{-0.0000040}$ \\
\ \ \ \ $\sigma_\textnormal{GP,24}$ [ppm] &  $\log U(10^{-4},10^2)$ &  $0.000100^{+0.000030}_{-0.0000017}$ \\
\ \ \ \ $\rho_\textnormal{GP,16}$ [d] &  $\log U(10^{-3},10^2)$ &  $73^{+16}_{-17}$ \\
\ \ \ \ $\rho_\textnormal{GP,17}$ [d] &  $\log U(10^{-3},10^2)$ &  $74^{+15}_{-18}$ \\
\ \ \ \ $\rho_\textnormal{GP,24}$ [d] &  $\log U(10^{-3},10^2)$ &  $76^{+14}_{-17}$ \\
\ \ \ \ $T_\textnormal{GP,16}$ [d] &  $\log U(10^{-3},10^2)$ &  $0.0010^{+0.000047}_{-0.000026}$ \\
\ \ \ \ $T_\textnormal{GP,17}$ [d] &  $\log U(10^{-3},10^2)$ &  $0.0011^{+0.000081}_{-0.000045}$ \\
\ \ \ \ $T_\textnormal{GP,24}$ [d] &  $\log U(10^{-3},10^2)$ &  $0.0010^{+0.000034}_{-0.000019}$ \\
\multicolumn{3}{l}{\textit{Ground-based photometry instrumental parameters}} \\
\ \ \ \ $q_{1,OAA}$ &  $U(0,1)$ &  $0.41^{+0.27}_{-0.24}$ \\
\ \ \ \ $q_{1,OMC}$ &  $U(0,1)$ &  $0.55^{+0.27}_{-0.31}$ \\
\ \ \ \ $q_{1,RCO}$ &  $U(0,1)$ &  $0.50^{+0.30}_{-0.29}$ \\
\ \ \ \ $m_\textnormal{flux,OAA}$ [ppm] &  $U(0,10^5)$ &  $-2148^{+214}_{-216}$ \\
\ \ \ \ $m_\textnormal{flux,OMC}$ [ppm] &  $U(0,10^5)$ &  $7936^{+6471}_{-6203}$ \\
\ \ \ \ $m_\textnormal{flux,RCO}$ [ppm] &  $U(0,10^5)$ &  $-979^{+6296}_{-5945}$ \\
\ \ \ \ $\sigma_\textnormal{w,OAA}$ [ppm] &  $\log U(0.1,10^5)$ &  $3395^{+146}_{-140}$ \\
\ \ \ \ $\sigma_\textnormal{w,OMC}$ [ppm] &  $\log U(0.1,10^5)$ &  $3123^{+202}_{-190}$ \\
\ \ \ \ $\sigma_\textnormal{w,RCO}$ [ppm] &  $\log U(0.1,10^5)$ &  $4143^{+165}_{-158}$ \\
\ \ \ \ $\theta_\textnormal{OAA}$ [d] &  $U(-10,10)$ &  $0.00112^{+0.00022}_{-0.00021}$ \\
\ \ \ \ $\theta_\textnormal{OMC}$ [d] &  $U(-10,10)$ &  $-0.0036^{+0.015}_{-0.015}$ \\
\ \ \ \ $\theta_\textnormal{RCO}$ [d] &  $U(-10,10)$ &  $0.013^{+0.015}_{-0.015}$ \\
\ \ \ \ $\sigma_\textnormal{GP,OMC}$ [ppm] &  $\log U(10^{-4},10^2)$ & $0.117^{0.022}_{-0.012}$\\
\ \ \ \ $\rho_\textnormal{GP,OMC}$ [d] &  $\log U(10^{-3},10^2)$ &  $0.438^{+0.040}_{-0.059}$ \\
\ \ \ \ $\sigma_\textnormal{GP,RCO}$ [ppm] &  $\log U(10^{-4},10^2)$ & $0.117^{0.019}_{-0.011}$\\
\ \ \ \ $\rho_\textnormal{GP,RCO}$ [d] &  $\log U(10^{-3},10^2)$ &  $0.438^{+0.039}_{-0.051}$ \\
\multicolumn{3}{l}{\textit{Radial-velocity instrumental/activity parameters}} \\
\ \ \ \ $\mu_{\rm CARMENES}$ [m\,s$^{-1}$] &  $U(-100,100)$ &  $0.1^{+4.7}_{-4.0}$ \\
\ \ \ \ $\sigma_{w,\rm CARMENES}$ [m\,s$^{-1}$] &  $\log U(0.01,100)$ &  $1.85^{+0.50}_{-0.49}$ \\
\ \ \ \ $\sigma_\textnormal{GP,\rm CARMENES}$ [m\,s$^{-1}$] &  $\log U(0.01,10^2)$ & $5.8^{+7.3}_{-2.5}$\\
\ \ \ \ $\alpha_\textnormal{GP,\rm CARMENES}$ (d$^{-1}$) &  $\log U(10^{-10},1)$ &  $0.000011^{+0.000433}_{-0.000011}$ \\
\ \ \ \ $\Gamma_\textnormal{GP,\rm CARMENES}$ &  $\log U(0.01, 100)$ &  $0.5^{+2.8}_{-0.5}$ \\
\ \ \ \ $P_\textnormal{GP,\rm Rot}$ &  $U(20, 350)$ &  $237^{+67}_{-103}$ 
\enddata
\tablenotetext{a}{These parametrize the quadratic limb-darkening law using the transformations in 
\cite{Kipping:LDs}}
\end{deluxetable*}

\begin{deluxetable}{lcl}
\tablecaption{Derived properties for TOI-1759~b.\label{tab:derivprops}}
\tablecolumns{3}
%\tablenum{2}
\tablewidth{0pt}
\tablehead{
\colhead{Parameter} &
\colhead{Posterior} & 
\colhead{Description}\\
}
\startdata
$R_p$ [$R_\oplus$] & $3.14 \pm 0.10$ & Planetary radius\\
$M_p$ [$M_\oplus$] & $10.8 \pm 1.5$ & Planetary mass\\
$i$ [deg] & $89.72 \pm 0.13$ & Orbital inclination\\
$T_{14}$ [hours] & $3.23 \pm 0.13$ & Transit duration\\
$a$ [au] & $0.1177 \pm 0.0038$ & Semi-major axis of the orbit\\
$T_{eq, 0}$ [K] & $443 \pm 7$ & Equilibrium temperature$^{a}$ \\
& & (assuming 0 albedo)\\
$T_{eq, 0.3}$ [K] & $405 \pm 6$ & Equilibrium temperature$^{a}$ \\
& & (assuming 0.3 albedo)\\
$S_p$ [$S_\oplus$] & $6.39 \pm 0.41$ & Stellar irradiation \\
& & on the planet\\
$\rho_p$ [g\,cm$^{-3}$] & $1.91 \pm 0.32$ & Planetary bulk density\\
$g_p$ [m\,s$^{-2}$] & $10.7 \pm 1.6$ & Planetary surface gravity\\
\enddata
\tablenotetext{a}{This assumes perfect energy redistribution.}
\end{deluxetable}

The resulting best-fit models and corresponding credibility bands are presented in 
Figure \ref{fig:tess-transits} for the \textit{TESS} photometry, in Figure \ref{fig:gb-transits} for 
the ground-based photometry and in Figure \ref{fig:carmenes-rvs} for the radial-velocities. The 
constraints from our radial-velocity follow-up allowed us to obtain 
a precise {{($>5\sigma$ above 0)}} measurement of the semi-amplitude imprinted by TOI-1759~b on its star of 
$K=3.64^{+0.50}_{-0.51}$\,m\,s$^{-1}$, which is an over 7-sigma detection of the semi-amplitude. Joining the 
derived transit and radial-velocity parameters, along with the stellar properties 
presented in Table \ref{tab:stprops}, we derive the fundamental parameters of TOI-1759~b in 
Table \ref{tab:derivprops}. As can be observed, TOI-1759~b is a relatively cool (443 K 
equilibrium temperature assuming 0 albedo) 
sub-Neptune-sized exoplanet ($R_p = 3.14 \pm 0.10 \,R_\oplus$). Coupling these numbers with 
our estimated mass of $M_p = 10.8 \pm 1.5 M_\oplus$, we derive a planetary bulk density ($\rho_p = 1.91\pm 0.32$ g\,cm$^{-3}$) 
and gravity ($g_p = 10.7\pm1.6$\, m\,s$^{-2}$) which are strikingly similar to those of Neptune (1.64 g\,cm$^{-3}$ and 
11.15\,m\,s$^{-2}$, respectively). We discuss the properties of TOI-1759~b in context of other 
discovered systems in the next section.

\section{Discussion} 
\label{sec:dis}

\begin{figure*}
\includegraphics[width=2.15\columnwidth]{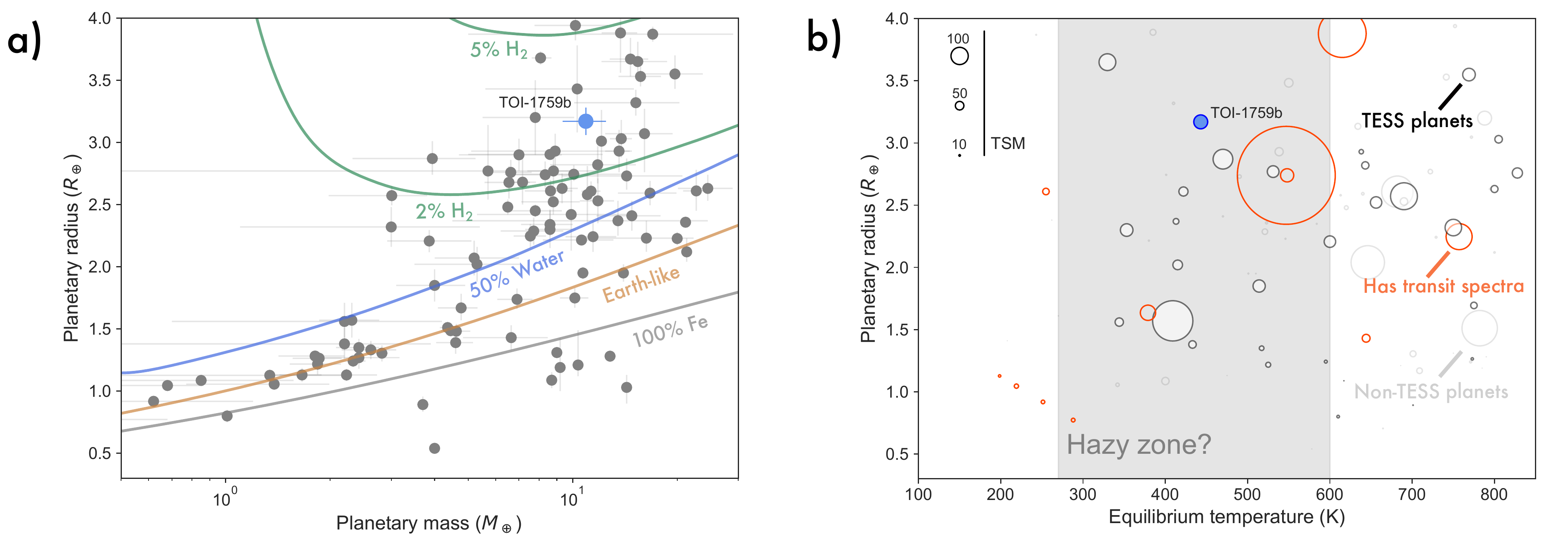}
\caption{\textbf{Properties of TOI-1759~b as compared to previously small, known exoplanets with measured mass and radius}. \textbf{a)} Mass-radius diagram of all known small ($R_p < 4\,R_\oplus$) exoplanets with $T_{\textnormal{eq}} < 1000$ K (grey) 
and TOI-1759~b (blue). The mass-radius models are from \cite{zeng16} and show 100\% Fe (grey), an Earth-like composition (32.5\% Fe, 67.5\% MgSiO$_3$ by mass, brown), a 50\% 
Earth-like and 50\% water-by-mass composition (blue) and Earth-like composition 
models with 2\% and 5\% H$_2$ on top of it (green). Water-rich models with 2\% and 5\% H$_2$ on 
top (i.e., Neptune-like compositions) are not shown, but would be indistinguishable with the 
green models in this panel given the size of the errorbars. \textbf{b)} Equilibrium temperature (assuming zero albedo) versus radius diagram for the same cuts made for 
panel a), showing the location of \textit{TESS} planets (white markers), non-\textit{TESS} planets (white transparent 
markers), planets that have previously been characterized via transmission 
spectroscopy with HST/WFC3 (red markers) and TOI-1759~b (blue). The size of the markers represents the value of the transmission spectroscopy metric 
\citep[TSM][]{tsm}; grey band shows the proposed region of hazy exoplanets 
by \cite[][see text for discussion]{Yu2021}. }
\label{fig:warm-worlds}
\end{figure*}

In order to put TOI-1759~b in context with the known sample of small exoplanets, we query 
the properties of all such exoplanets that (a) have 
both a measured mass and radius, (b) are smaller than $R_p < 4\,R_\oplus$ and (c) have 
equilibrium temperatures cooler than $1000$\,K from the NASA Exoplanet Archive \citep{nexsci} 
--- i.e., a cut similar to that presented in \cite{guo2020}, but updated with the latest 
exoplanetary systems as of July 23, 2021\footnote{More recent queries, along with the same 
plots shown here, can be generated using the scripts in this 
repository: \url{https://github.com/nespinoza/warm-worlds}}. In Figure~\ref{fig:warm-worlds}, we show the location of 
TOI-1759~b in both the planetary mass versus radius plane 
and the equilibrium temperature versus planetary radius plane.

In terms of its mass and radius, Figure \ref{fig:warm-worlds}a shows that TOI-1759~b is 
consistent with having a $2-5\%$ H$_2$ envelope, and an interior composition ranging from being an Earth-like one to being a scaled-down 
version of Neptune. Figure \ref{fig:warm-worlds}b, on the other hand, shows how TOI-1759~b adds 
up to the increasing number of small ($R_p<4\,R_\oplus$) worlds with measured mass and radius at relatively low equilibrium temperatures. In particular, TOI-1759~b falls on the very interesting region where the work of \cite{Yu2021} recently 
proposed exoplanet atmospheres to be hazy due to the lack of haze-removal processes at 
temperatures between about 300-600 K. It is interesting to note that TOI-1759~b falls \textit{exactly} at the equilibrium temperature where \cite{Yu2021} predict 
the haziest exoplanets should be ($T_{eq,0.3}\sim 400$ K, where $T_{eq,0.3}$ means an 
equilibrium temperature calculated assuming an albedo of 0.3; see Table 
\ref{tab:derivprops}). The proposed trend presented in that work 
seems to be in line with observed transmission spectra for planets hotter than about 
500 K \citep{ck:2017}. For example, GJ~1214~b \citep[550 K,][biggest red circle in Figure \ref{fig:warm-worlds}a]{gj1214} shows a significantly muted water feature, whereas 
HAT-P-11~b \citep[750 K,][not shown in Figure \ref{fig:warm-worlds}a as $R_p=4.3\,R_\oplus$ 
for this exoplanet]{hatp11} has a $3-$scale height water amplitude in its transmission 
spectrum. However, the hypothesis is harder to test 
for temperate exoplanets ($< 500$\,K), 
as good 
targets for atmospheric characterization have 
remained scarce, in particular for small ($R_p < 4\,R_\oplus$) planets.

%On another note, understanding abiogenic methane production and destruction processes in the exoplanet atmospheres paves the way for investigating bio-signatures in the near future \citep[e.g.][]{green2021call}. While methane is expected to be present in the atmosphere of exoplanets over a wide range of parameter space and under thermochemical assumption \citep[e.g.][]{molaverdikhani2019colda,bezard2020}, there has been only one robust methane detection \citep[][]{giacobbe2021five}. This suggests atmospheric processes, such as internal heating \citep[e.g.][]{benneke2019sub,fortney2020beyond}, strong vertical mixing \citep[e.g.][]{venot2014influence,molaverdikhani2019coldb}, and cloud blanket effect \citep[e.g.][]{molaverdikhani2020role} might be in action to cause such methane depletion. But suitable targets for atmospheric spectroscopy are needed to examine these hypotheses for the temperate exoplanets regime. -> I removed this part. It's a bit off-topic (or at least, not in line with the punch line of "this is a good target for atmospheric characterization".

TOI-1759~b is among the best temperate targets to perform transmission spectroscopy based on its 
Transmission Spectroscopy Metric \citep[TSM][]{tsm}. Following the work of \cite{tsm}, we 
estimate a TSM of $81 \pm 14$, which puts it among the top five targets for atmospheric 
characterization to date at equilibrium temperatures lower than 500 K, 
together with L 98-59~d \citep[TSM of 233;][]{cloutier19, kostov19, pidhorodetska21}, 
TOI-178~g \citep[TSM of 114;][]{leleu21}, 
TOI-1231~b \citep[TSM of 97;][]{burt2021} and LHS~1140~b 
\citep[TSM of 89;][]{dittman17, ment19} --- the latter 
having actually been recently characterized by HST/WFC3 \citep{edwards21}, reporting weak evidence for water absorption in its planetary atmosphere.

For a quantitative assessment of TOI-1759 b’s atmospheric characterization with JWST, we investigated a suite of atmospheric scenarios and calculated their JWST synthetic spectra using the photo-chemical model ChemKM \citep[][]{molaverdikhani2019coldb,molaverdikhani2020understanding}, the radiative transfer model petitRADTRANS \citep[][]{molliere2019petitradtrans,molliere2020retrieving}, and \texttt{ExoTETHyS} \citep{morello2021} for uncertainty estimations.

\begin{figure}
\includegraphics[width=1\columnwidth]{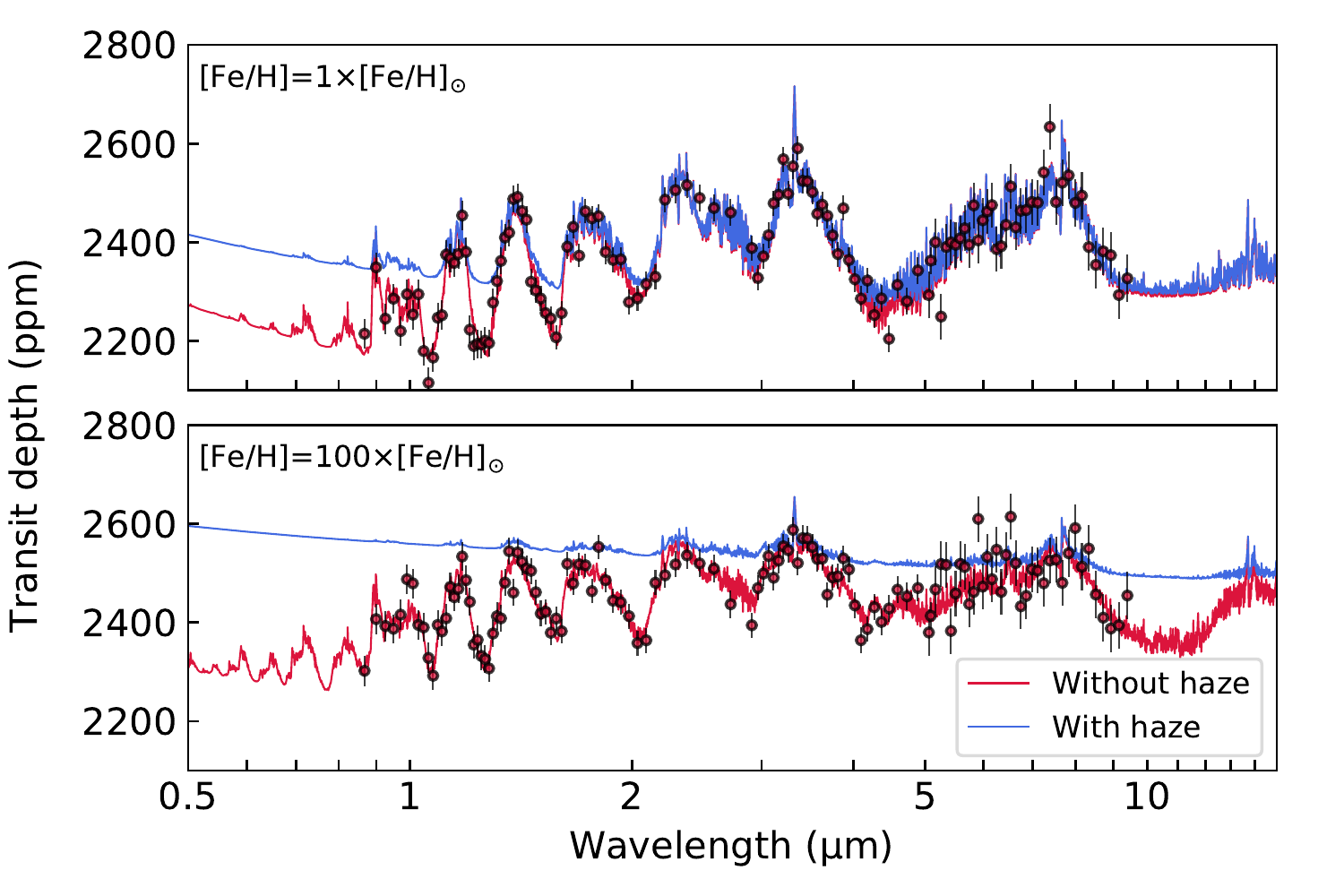}
\caption{\textbf{Synthetic JWST transmission atmospheric spectra of TOI-1759~b}. Top : Fiducial models with solar
abundance, Bottom : Enhanced metallicity by a factor of 100, by considering haze opacity (solid blue lines) and without haze opacity (solid red lines). Estimated uncertainties are shown
for the observation of one transit with JWST NIRISS-SOSS, NIRSpec-G395M, and MIRI-LRS configurations.}
\label{fig:synth_spectra}
\end{figure}

Assuming an isothermal atmosphere with a temperature of 400 K and a constant vertical mixing of $K_{zz}$=$10^6$\, cm$^2$ s$^{-1}$ results in persisting water and methane features in the transmission spectra of TOI-1759~b, see \autoref{fig:synth_spectra}. But such atmospheric features are expected to be suppressed in a high-metallicity atmosphere, see bottom panel in \autoref{fig:synth_spectra}. Considering haze in the atmosphere of TOI-1759~b further mutes the features and hence a hazy, high-metallicity atmosphere is expected to show a nearly flat transmission spectrum, see the blue line in \autoref{fig:synth_spectra} bottom panel.

The \texttt{PandExo} package \citep{batalha2017} was used to determine the best configurations to observe with the NIRISS SOSS (0.6-2.8 $\mu$m), NIRSpec G395M (2.88-5.20 $\mu$m) and MIRI LRS (5-12 $\mu$m) instrumental modes. Then we used \texttt{ExoTETHyS}  to compute the simulated spectra. The wavelength bins were specifically determined to have similar counts, leading to nearly uniform error bars per spectral point. Note that the minimal error bars output by \texttt{ExoTETHyS} have been multiplied by the reciprocal of the square root of the observing efficiency and a conservative factor 1.2 that accounts for correlated noise. The resulting error bars are equal to or slightly larger than those obtained with \texttt{PandExo} for the same wavelength bins. In particular, the spectral error bars estimated for just one transit observation per instrument configuration are 25-30 ppm at wavelengths $<$5 $\mu$m, and 45-50 ppm at wavelengths $>$5 $\mu$m, with several points to sample each molecular feature as shown in \autoref{fig:synth_spectra}). Comparing these uncertainties with the expected water and methane features of $\sim$200~ppm significance suggests the possibility of differentiating these scenarios during one transit only.

TOI-1759~b, along with TOI-178~g ($R_p = 2.87\,R_\oplus$, $T_{eq,0} = 470$\,K), TOI-1231~b 
($R_p = 3.65\,R_\oplus$, $T_{eq,0} = 330$\,K) and the (now) iconic K2-18~b 
\citep[$R_p = 2.61\,R_\oplus$, $T_{eq,0} = 255$\,K,][]{b19, t19} form an excellent 
sample of sub-Neptunes to perform atmospheric characterization via 
transmission spectroscopy at equilibrium temperatures below 500 K. %as sparsely as they are for larger equilibrium temperatures; 
A sample that can be used to put the prediction of both proposed haze removal \citep{Yu2021} and methane removal \citep[][]{molaverdikhani2020role} processes to the test.

% \subsection{Is there a TOI-1759~c?}

%While in Section \ref{sec:ana} we performed an analysis assuming the best model to use for our global analysis 
%was the one %in which the radial-velocities were best explained by 
%with a planetary signal plus a Gaussian Process (based on 
%our radial-velocity analysis presented in Section \ref{ssec:rvana}), the 
%analysis in Section \ref{ssec:rvana} also revealed that, formally, a 
%two-planet model was indistinguishable from a one-planet-plus-GP model (see Table \ref{tab:logZ}). A two-Keplerian global fit of the data gives rise to 
%the same properties as those reported in Section \ref{sec:glob} for TOI-1759~b; 
%however, from our radial-velocity analysis of the activity indicators in 
%Section \ref{ssec:rvana}, it seems likely that while the 80-day periodicity 
%observed in the data is consistent with stellar activity, a $>200$-day 
%periodicity is still present in the data, with a semi-amplitude of about 
%4\,m\,s$^{-1}$. Given the properties of the star, this would imply a Saturn-mass 
%companion on a wide orbit with an equilibrium temperature of about 200 K. 
%The mass ratio between a Saturn-mass planet and the star would be 
%about $q = 10^{-4}$, which puts it comfortably among expected mass ratios given the 
%$M_\star = 0.6\,M_\odot$ stellar host \citep{morales19}. More observations with a 
%longer time-baseline than the one presented here --- hopefully spanning several 
%years --- are needed in order to unveil if indeed this long-term 
%periodicity is real, along with the signal's true nature.

\section{Conclusions}
\label{sec:conclusions}

We have presented the discovery and characterization of the transiting exoplanet 
TOI-1759~b, a sub-Neptune ($R_p=3.14 \pm 0.10 \,R_\oplus$, $M_p=10.8 \pm 1.5\,M_\oplus$) exoplanet 
on a 18.85-day orbit around an M~dwarf star. The 
initial identification of the target was made thanks to precise \textit{TESS} photometry, 
which unveiled three transits of the exoplanet in three different sectors with an 
ambigous period being consistent with both, a 18.85 and a 37.7-day period exoplanet. 
Thanks to ground-based photometric follow-up from different observatories, 
high-resolution spatial imaging and precise radial-velocities from the CARMENES 
high-resolution spectrograph, we were able to not only confirm TOI-1759~b as a {bona fide} transiting exoplanet and precisely measure its mass, but also constrain its true period to 
be $18.85019 \pm 0.00013$\,d. 

TOI-1759~b adds to the growing number of temperate ($T_\textnormal{eq} < 500$\,K) 
exoplanets, and is a particularly promising target to perform atmospheric characterization on. Its equilibrium temperature ($T_\textnormal{eq,0.3} = 405 \pm 6$\,K) puts it exactly where the work of \cite{Yu2021} predicts the haziest exoplanets to be, and thus 
provides an exciting system in which to test this proposal. In addition, our 6-month 
radial-velocity campaign revealed an 80-day periodicity in 
the data most likely arising from 
stellar activity, and a possible longer-term periodicity with a period $> 200$\,d. 
{{The current baseline of our CARMENES observations is insufficient 
to unveil the true nature of this latter long-period signal. However, a campaign spanning a longer baseline 
is needed in order to reveal the exact source and periodicity of this signal. }}

\acknowledgements
%
%This work has made use of data from the European Space Agency (ESA) mission {\em Gaia} (\url{https://www.cosmos.esa.int/gaia}), processed by the {\em Gaia} Data Processing and Analysis Consortium (DPAC, \url{ https://www.cosmos.esa.int/web/gaia/dpac/consortium}). 
%Funding for the DPAC has been provided by national institutions, in particular the institutions participating in the {\em Gaia} MultilateralAgreement.
 CARMENES is an instrument at the Centro Astron\'omico Hispano-Alem\'an (CAHA) at Calar Alto (Almer\'{\i}a, Spain), operated jointly by the Junta de Andaluc\'ia and the Instituto de Astrof\'isica de Andaluc\'ia (CSIC).
 CARMENES was funded by the Max-Planck-Gesellschaft (MPG),  the Consejo Superior de Investigaciones Cient\'{\i}ficas (CSIC), the Ministerio de Econom\'ia y Competitividad (MINECO) and the European Regional Development Fund (ERDF) through projects FICTS-2011-02, ICTS-2017-07-CAHA-4, and CAHA16-CE-3978, and the members of the CARMENES Consortium  (Max-Planck-Institut f\"ur Astronomie, Instituto de Astrof\'{\i}sica de Andaluc\'{\i}a, Landessternwarte K\"onigstuhl, Institut de Ci\`encies de l'Espai, Institut f\"ur Astrophysik G\"ottingen, Universidad Complutense de Madrid, Th\"uringer Landessternwarte Tautenburg, Instituto de Astrof\'{\i}sica de Canarias,  Hamburger Sternwarte, Centro de Astrobiolog\'{\i}a and Centro Astron\'omico Hispano-Alem\'an), with additional contributions by the MINECO,  the Deutsche Forschungsgemeinschaft through the Major Research Instrumentation Programme and Research Unit FOR2544 ``Blue Planets around Red Stars'', the Klaus Tschira Stiftung, the states of Baden-W\"urttemberg and Niedersachsen, and by the Junta de Andaluc\'{\i}a. 
 This work was based on data from the CARMENES data archive at CAB (CSIC-INTA).
 We acknowledge financial support from the Agencia Estatal de Investigaci\'on of the Ministerio de Ciencia, Innovaci\'on y Universidades and the ERDF through projects 
 PID2019-109522GB-C5[1:4],	% CAB+IAA+IAC+UCM
 PGC2018-098153-B-C33,	% ICE
 AYA2018-84089, 	% CAB SVO
 PID2019-107061GB-C64, PID2019-110689RB-100 % Eloy
  %ESP2017-87676-C5-1-R, % CAB PLATO  
AYA2016-79425-C3-1/2/3-P and BES-2017-080769,
and the Centre of Excellence ``Severo Ochoa'' and ``Mar\'ia de Maeztu'' awards to the Instituto de Astrof\'isica de Canarias (CEX2019-000920-S), Instituto de Astrof\'isica de Andaluc\'ia (SEV-2017-0709), and Centro de Astrobiolog\'ia (MDM-2017-0737), NASA (NNX17AG24G), and the Generalitat de Catalunya/CERCA programme. {{Data were partly collected with the 90-cm telescope at the Sierra Nevada 
Observatory (SNO) operated by the Instituto de Astrof\'\i fica de 
Andaluc\'\i a (IAA, CSIC). We acknowledge the telescope operators from the 
Sierra Nevada Observatory for their support. G. M. has received funding from the European Union’s Horizon 2020
research and innovation programme under the Marie Sklodowska-Curie grant
agreement No 895525.}} {{This research has made use of the NASA Exoplanet Archive, which is operated by the California Institute of Technology, under contract with the National Aeronautics and Space Administration under the Exoplanet Exploration Program.}} We acknowledge the use of public TESS data from pipelines at the TESS Science Office and at the TESS Science Processing Operations Center. Resources supporting this work were provided by the NASA High-End Computing (HEC) Program through the NASA Advanced Supercomputing (NAS) Division at Ames Research Center for the production of the SPOC data products. 
The authors wish to recognize and acknowledge the very significant cultural role and reverence that the summit of Maunakea has always had within the indigenous Hawaiian community.  We are most fortunate to have the opportunity to conduct observations from this mountain. 

\vspace{5mm}
\facilities{\textit{TESS}, CARMENES/3.5-m Calar Alto telescope, TJO, {{SNO}}, AAM, MONTSEC, Keck telescope, Gemini-North telescope}.

\software{          
          \textsf{radvel} \citep{fulton:2018},
          \textsf{batman} \citep{kreidberg:2015}, 
          \textsf{juliet} \citep{espinoza:juliet}, 
          \textsf{astroimagej}
          \citep{collins17},
          \textsf{tpfplotter}
          \citep{Aller2020},
          \textsf{dynesty}
          \citep{dynesty},
          \textsf{celerite}
          \citep{celerite}.
          }
    
\appendix

\section{Radial-velocity data}

Our full CARMENES dataset for the VIS channel is presented in Table \ref{tab:rvs}, along with the corresponding activity 
indicators at each epoch.

\begin{deluxetable}{lrrrrrrrrrrr}
\tablewidth{0pc} 
\tablecaption{
    Radial velocity measurements for the star along with activity indicators at each epoch. \label{tab:rvs}
}
\tablehead{
    \colhead{Name} &
    \colhead{BJD} & 
    \colhead{RV} & 
    \colhead{\ensuremath{\sigma_{\rm RV}}} & 
    \colhead{CRX} & 
    \colhead{dLW} & 
    \colhead{BIS} & 
    \colhead{\ensuremath{\rm{H}_{\alpha}}} & 
    \colhead{Ca II IRT a} & 
    \colhead{Instrument} &
    \colhead{S/N}\\ 
    \colhead{} & 
    \colhead{-2450000} &  
    \colhead{(m s$^{-1}$)} &  
    \colhead{(m s$^{-1}$)} & 
    \colhead{(m s$^{-1}$ Np$^{-1}$)} & 
    \colhead{($10^{3}$ m$^{2}$ s$^{-2}$)} &
    \colhead{(km s$^{-1}$)} &
    \colhead{(km s$^{-1}$)} &
    \colhead{(km s$^{-1}$)} &
    \colhead{} &
    \colhead{} \\ 
} 
\startdata 
TOI-1759 & 9054.56851 & 2.33 & 2.51 & 49.44 & 1.48 & -0.0774 & 0.074 & 0.0094 & CARMENES-VIS & 119.4  \\ 
TOI-1759 & 9067.60481 & 3.32 & 2.32 & 43.58 & -0.04 & -0.0646 & 0.082 & 0.0128 & CARMENES-VIS & 97.6  \\ 
TOI-1759 & 9068.57556 & 1.86 & 1.86 & 29.70 & -11.16 & -0.0776 & 0.074 & 0.0290 & CARMENES-VIS & 94.3  \\ 
TOI-1759 & 9069.59957 & 0.86 & 1.78 & 21.57 & -5.31 & -0.0635 & 0.054 & 0.0154 & CARMENES-VIS & 101.8  \\ 
TOI-1759 & 9070.55690 & 2.17 & 1.85 & 7.88 & -5.34 & -0.0764 & 0.076 & 0.0237 & CARMENES-VIS & 111.2  \\ 
TOI-1759 & 9076.57116 & 6.20 & 2.55 & 15.86 & -22.86 & -0.0763 & 0.078 & 0.0304 & CARMENES-VIS & 80.6  \\ 
TOI-1759 & 9078.60935 & 8.44 & 2.01 & 18.62 & -11.90 & -0.0719 & 0.068 & 0.0068 & CARMENES-VIS & 103.2  \\ 
TOI-1759 & 9079.59579 & 3.59 & 2.33 & 36.88 & -16.68 & -0.0677 & 0.079 & 0.0154 & CARMENES-VIS & 104.2  \\ 
TOI-1759 & 9081.58773 & 6.10 & 2.47 & -16.45 & -19.67 & -0.0604 & 0.079 & 0.0099 & CARMENES-VIS & 81.8  \\ 
TOI-1759 & 9084.55634 & 0.35 & 1.48 & -10.41 & 2.25 & -0.0560 & 0.093 & 0.0120 & CARMENES-VIS & 103.7  \\ 
TOI-1759 & 9087.59796 & 1.37 & 2.12 & 6.67 & -1.17 & -0.0517 & 0.069 & 0.0071 & CARMENES-VIS & 104.8  \\ 
TOI-1759 & 9089.53629 & -1.43 & 1.95 & -8.83 & -5.14 & -0.0600 & 0.071 & 0.0110 & CARMENES-VIS & 109.4  \\ 
TOI-1759 & 9090.55842 & -3.51 & 2.75 & -8.03 & -14.72 & -0.0490 & 0.069 & 0.0033 & CARMENES-VIS & 76.3  \\ 
TOI-1759 & 9091.54399 & -8.08 & 3.58 & -32.14 & -6.07 & -0.0609 & 0.049 & 0.0216 & CARMENES-VIS & 48.5  \\ 
&&&&& $\cdot \cdot \cdot$ &&&&&\\
TOI-1759 & 9173.30701 & -3.00 & 2.43 & -12.31 & 12.00 & -0.0291 & 0.052 & 0.0104 & CARMENES-VIS & 93.4  \\ 
TOI-1759 & 9174.31626 & -2.65 & 1.89 & -11.97 & 15.74 & -0.0118 & 0.065 & -0.0023 & CARMENES-VIS & 90.7  \\ 
TOI-1759 & 9175.35015 & 6.53 & 2.21 & -32.46 & 7.19 & -0.0132 & 0.102 & 0.0127 & CARMENES-VIS & 77.8  \\ 
TOI-1759 & 9176.32926 & -10.27 & 2.28 & -14.38 & 20.12 & -0.0227 & 0.064 & 0.0020 & CARMENES-VIS & 114.3  \\ 
TOI-1759 & 9177.32923 & -1.93 & 1.78 & -35.19 & 13.28 & -0.0312 & 0.066 & 0.0014 & CARMENES-VIS & 111.0  \\ 
TOI-1759 & 9178.28239 & -11.01 & 1.90 & -15.77 & 7.78 & -0.0338 & 0.086 & 0.0099 & CARMENES-VIS & 95.8  \\ 
TOI-1759 & 9183.29700 & -10.37 & 2.20 & -31.60 & -2.18 & -0.0157 & 0.071 & 0.0123 & CARMENES-VIS & 97.3  \\ 
TOI-1759 & 9186.33693 & -9.05 & 4.74 & -45.55 & -22.24 & -0.0773 & 0.037 & 0.0161 & CARMENES-VIS & 27.1  \\ 
TOI-1759 & 9187.30129 & -1.25 & 3.31 & -45.82 & -24.91 & -0.0299 & 0.103 & -0.0038 & CARMENES-VIS & 50.3  \\ 
TOI-1759 & 9193.27483 & 4.02 & 3.18 & 0.08 & -0.95 & -0.0247 & 0.084 & -0.0052 & CARMENES-VIS & 64.2  \\ 
TOI-1759 & 9196.26135 & -8.36 & 2.97 & -9.94 & 7.30 & -0.0186 & 0.032 & -0.0079 & CARMENES-VIS & 72.1  \\ 
TOI-1759 & 9197.34852 & -4.98 & 3.64 & -22.89 & -3.08 & -0.0198 & 0.085 & 0.0218 & CARMENES-VIS & 57.4  \\ 
TOI-1759 & 9209.39053 & 6.56 & 6.36 & 7.17 & -27.90 & -0.0371 & 0.026 & 0.0052 & CARMENES-VIS & 25.7  \\ 
TOI-1759 & 9216.25949 & 1.84 & 5.71 & -48.81 & 3.21 & -0.0156 & 0.125 & 0.0121 & CARMENES-VIS & 38.2  \\ 
TOI-1759 & 9218.29595 & -5.22 & 2.45 & -15.32 & -17.68 & -0.0126 & 0.082 & 0.0091 & CARMENES-VIS & 61.6  \\ 
TOI-1759 & 9219.28687 & -4.05 & 3.34 & -61.08 & -22.66 & -0.0350 & 0.055 & 0.0109 & CARMENES-VIS & 47.8  \\ 
TOI-1759 & 9231.28008 & 1.00 & 2.57 & -4.70 & 3.51 & -0.0059 & 0.052 & -0.0024 & CARMENES-VIS & 92.8  \\ 
TOI-1759 & 9232.28137 & 2.54 & 2.77 & -4.43 & 10.19 & -0.0298 & 0.052 & -0.0001 & CARMENES-VIS & 96.4  \\ 
\enddata
\tablecomments{
    Signal-to-noise ratio (S/N) for CARMENES-VIS data corresponds to the S/N at order 36 (at about $840$ nm). 
    A sample of the full radial-velocity dataset and activity indicators are shown here. The entirety of this 
    table is available in a machine-readable form in the online journal.
}
\end{deluxetable}

\section{Photometric data}

Our full photometric dataset targeting transits of TOI-1759 is presented in Table \ref{tab:lcs}. The long-term photometry is 
presented in Table \ref{tab:ltphot}.

\begin{deluxetable}{lrrrrrr}
\tablewidth{0pc} 
\tablecaption{
    Photometric transits of the planet. \label{tab:lcs}
}
\tablehead{
    \colhead{Name} &
    \colhead{BJD} & 
    \colhead{Relative flux} & 
    \colhead{Error} & 
    \colhead{Instrument} &
    \colhead{Standarized airmass}\\ 
    \colhead{} & 
    \colhead{-2450000} &  
    \colhead{} &  
    \colhead{} & 
    \colhead{} & 
    \colhead{} \\ 
} 
\startdata 
TOI-1759 & 8738.65456 & 1.000458 & 0.001213 & TESS - Sector 16 & --- \\ 
TOI-1759 & 8738.65594 & 0.999531 & 0.001211 & TESS - Sector 16 & --- \\ 
TOI-1759 & 8738.65733 & 1.000397 & 0.001213 & TESS - Sector 16 & --- \\ 
TOI-1759 & 8738.65872 & 1.001858 & 0.001212 & TESS - Sector 16 & --- \\ 
TOI-1759 & 8738.66011 & 0.999702 & 0.001212 & TESS - Sector 16 & --- \\ 
TOI-1759 & 8738.66150 & 0.999855 & 0.001212 & TESS - Sector 16 & --- \\ 
TOI-1759 & 8738.66289 & 0.999816 & 0.001210 & TESS - Sector 16 & --- \\ 
TOI-1759 & 8738.66428 & 0.999837 & 0.001209 & TESS - Sector 16 & --- \\ 
TOI-1759 & 8738.66567 & 0.998040 & 0.001210 & TESS - Sector 16 & --- \\ 
TOI-1759 & 8738.66706 & 0.998851 & 0.001212 & TESS - Sector 16 & --- \\ 
TOI-1759 & 8738.66844 & 0.999615 & 0.001213 & TESS - Sector 16 & --- \\ 
&& $\cdot \cdot \cdot$ &&&\\
TOI-1759 & 8990.54955 & 0.995134 & 0.001107 & Albanya-0.4m (OAA-Ic) & -1.372 \\ 
TOI-1759 & 8990.55022 & 1.000704 & 0.001110 & Albanya-0.4m (OAA-Ic) & -1.380 \\ 
TOI-1759 & 8990.55090 & 0.998573 & 0.001110 & Albanya-0.4m (OAA-Ic) & -1.388 \\ 
TOI-1759 & 8990.55159 & 0.997527 & 0.001110 & Albanya-0.4m (OAA-Ic) & -1.396 \\ 
TOI-1759 & 8990.55225 & 0.999177 & 0.001110 & Albanya-0.4m (OAA-Ic) & -1.403 \\ 
TOI-1759 & 8990.55293 & 1.001281 & 0.001110 & Albanya-0.4m (OAA-Ic) & -1.411 \\ 
TOI-1759 & 8990.55360 & 0.998378 & 0.001107 & Albanya-0.4m (OAA-Ic) & -1.419 \\ 
TOI-1759 & 8990.55428 & 0.998890 & 0.001113 & Albanya-0.4m (OAA-Ic) & -1.426 \\ 
TOI-1759 & 8990.55496 & 0.998735 & 0.001113 & Albanya-0.4m (OAA-Ic) & -1.434 \\ 
TOI-1759 & 8990.55563 & 0.997969 & 0.001113 & Albanya-0.4m (OAA-Ic) & -1.442 \\ 
\enddata
\tablecomments{
    The Standarized Airmass column corresponds to the airmass values substracted by their mean and divided by their standard deviation. 
    A sample of the full photometric dataset is shown here. The entirety of this table is available in a 
    machine-readable form in the online journal or the source file used to generate this compiled PDF version. 
}
\end{deluxetable}

\begin{deluxetable}{lrrrrr}
\tablewidth{0pc} 
\tablecaption{
    Long-term photometric measurements of TOI-1759. \label{tab:ltphot}
}
\tablehead{
    \colhead{Name} &
    \colhead{BJD} & 
    \colhead{Relative flux} & 
    \colhead{Error} & 
    \colhead{Instrument} \\ 
    \colhead{} & 
    \colhead{-2450000} &  
    \colhead{} &  
    \colhead{} & 
    \colhead{} \\ 
} 
\startdata 
TOI-1759 & 9010.61701 & 0.993732 & 0.000714 & Joan-Oro-0.8m (TJO) \\ 
TOI-1759 & 9010.61758 & 0.993128 & 0.000714 & Joan-Oro-0.8m (TJO) \\ 
TOI-1759 & 9010.61813 & 0.991751 & 0.000714 & Joan-Oro-0.8m (TJO) \\ 
TOI-1759 & 9010.61868 & 0.992962 & 0.000714 & Joan-Oro-0.8m (TJO) \\ 
TOI-1759 & 9010.61923 & 0.992559 & 0.000714 & Joan-Oro-0.8m (TJO) \\ 
TOI-1759 & 9010.61978 & 0.990158 & 0.000714 & Joan-Oro-0.8m (TJO) \\ 
TOI-1759 & 9010.62033 & 0.990918 & 0.000714 & Joan-Oro-0.8m (TJO) \\ 
TOI-1759 & 9010.62089 & 0.991625 & 0.000714 & Joan-Oro-0.8m (TJO) \\ 
TOI-1759 & 9010.62145 & 0.992897 & 0.000719 & Joan-Oro-0.8m (TJO) \\ 
TOI-1759 & 9010.62200 & 0.991557 & 0.000719 & Joan-Oro-0.8m (TJO) \\ 
TOI-1759 & 9010.62255 & 0.991023 & 0.000719 & Joan-Oro-0.8m (TJO) \\ 
 &  & $\cdot \cdot \cdot$ & &  \\ 
TOI-1759 & 9457.65199 & 1.003121 & 0.001813 & Sierra-Nevada-0.9m (SNO-V) \\ 
TOI-1759 & 9457.65339 & 1.005637 & 0.001809 & Sierra-Nevada-0.9m (SNO-V) \\ 
TOI-1759 & 9457.65479 & 1.003229 & 0.001785 & Sierra-Nevada-0.9m (SNO-V) \\ 
TOI-1759 & 9457.65619 & 1.003976 & 0.001838 & Sierra-Nevada-0.9m (SNO-V) \\ 
TOI-1759 & 9457.65759 & 1.003824 & 0.001882 & Sierra-Nevada-0.9m (SNO-V) \\ 
TOI-1759 & 9457.65899 & 1.006960 & 0.001844 & Sierra-Nevada-0.9m (SNO-V) \\ 
TOI-1759 & 9457.66039 & 1.001677 & 0.001862 & Sierra-Nevada-0.9m (SNO-V) \\ 
TOI-1759 & 9457.66320 & 1.000984 & 0.001788 & Sierra-Nevada-0.9m (SNO-V) \\ 
TOI-1759 & 9457.66460 & 1.005191 & 0.001829 & Sierra-Nevada-0.9m (SNO-V) \\ 
TOI-1759 & 9457.66600 & 1.002202 & 0.001869 & Sierra-Nevada-0.9m (SNO-V) \\ 
\enddata
\tablecomments{
 A sample of the full photometric dataset is shown here. The entirety of this table is available in a machine-readable form in the online journal or the source file used to generate this compiled PDF version.
}
\end{deluxetable}

\bibliography{tessbib}
\end{document}